\documentclass[12pt]{article}

\usepackage[margin=2.5cm]{geometry}
\usepackage[T1]{fontenc}

\usepackage{amsmath}
\usepackage{
    accents, adjustbox, afterpage, amssymb, amsfonts, amsthm, array, booktabs, cancel, caption, color, comment, datetime, dcolumn, dsfont, eurosym, float, geometry, graphicx, lipsum, longtable, mathtools, mathbbol, multirow, pdflscape, pdfpages, ragged2e, rotating, setspace, subcaption, subfiles, textpos, threeparttable, soul, xcolor,
}

\usepackage[hang, flushmargin, bottom]{footmisc}

\usepackage{enumitem}
\usepackage{tikz}

\usepackage[authoryear]{natbib}
\setcitestyle{authoryear,aysep={},open={(},close={)}}
\bibpunct{(}{)}{;}{a}{,}{,}

\usepackage[hypertexnames=false,colorlinks]{hyperref}
\hypersetup{
    colorlinks,
    linkcolor={red!50!black},
    citecolor={blue!50!black},
    urlcolor={blue!50!black}
}

\usepackage{etoolbox}
\usepackage{xparse}
\NewDocumentCommand{\hyref}{m O{}O{}}{\hyperref[#1]{#2 \ref{#1}#3}}

\newdateformat{DDMonthYYYY}{%
  \THEDAY\, \monthname[\THEMONTH] \THEYEAR
  }

\theoremstyle{plain}
\newtheorem{theorem}{Theorem}
\newtheorem{proposition}{Proposition}
\newtheorem{lemma}{Lemma}

\newtheorem{corollary}{Corollary}

\theoremstyle{definition}
\newtheorem{definition}{Definition}

\newtheorem{remark}{Remark}

\newcommand*\diff{\mathop{}\!\mathrm{d}}

\usepackage{dsfont}
\usepackage{fouriernc}
\usepackage{charter}
\usepackage{libertine}

\newcommand\blfootnote[1]{
  \begingroup
  \renewcommand\thefootnote{}\footnote{#1}
  \addtocounter{footnote}{-1}
  \endgroup
}

\usepackage[ruled,vlined]{algorithm2e}
\usepackage{bookmark}

\usepackage[sf,sl,outermarks]{titlesec}
\newcommand{\addperiod}[1]{#1.}

\titleformat{\section}[block]
{\normalfont\Large\bfseries}{\thesection.}{.5em}{\Large\bfseries}
\titlespacing*{\section}{0pt}{*1.3}{*0.2}

\titleformat{\subsection}[block]
{\normalfont\large\bfseries}{\thesubsection.}{.5em}{\large\bfseries}
\titlespacing*{\subsection}{0pt}{*1}{*0}

\titleformat{\subsubsection}[runin]
{\normalfont\bfseries}{}{0em}{\normalsize\bfseries\addperiod}
\titlespacing*{\subsubsection}{0pt}{*1}{*1}

\titleformat{\paragraph}[runin]
{\normalfont\itshape}{}{0em}{\normalsize\itshape\addperiod}
\titlespacing*{\paragraph}{0pt}{*1}{*1}

\definecolor{mathematica1}{rgb}{0.368417, 0.506779, 0.709798}
\definecolor{mathematica2}{rgb}{0.880722, 0.611041, 0.142051}

\hypersetup{
    pdffitwindow=false,
    pdfstartview={XYZ null null 1.00},
    pdfnewwindow=true,
    colorlinks=true,
    linkcolor={red!50!black},
    citecolor={blue!50!black},
    urlcolor=black,
	linkbordercolor={white},
    pdfpagemode = UseNone,
}

\makeatletter
\DeclareRobustCommand\citepos													
  {\begingroup\def\NAT@nmfmt##1{{\NAT@up##1's}}%
   \NAT@swafalse\let\NAT@ctype\z@\NAT@partrue
   \@ifstar{\NAT@fulltrue\NAT@citetp}{\NAT@fullfalse\NAT@citetp}}

\pretocmd{\NAT@citex}{%
  \let\NAT@hyper@\NAT@hyper@citex
  \def\NAT@postnote{#2}%
  \setcounter{NAT@total@cites}{0}%
  \setcounter{NAT@count@cites}{0}%
  \forcsvlist{\stepcounter{NAT@total@cites}\@gobble}{#3}}{}{}
\newcounter{NAT@total@cites}
\newcounter{NAT@count@cites}
\def\NAT@postnote{}

\def\NAT@hyper@citex#1{
  \stepcounter{NAT@count@cites}%
  \hyper@natlinkstart{\@citeb\@extra@b@citeb}#1%
  \ifnumequal{\value{NAT@count@cites}}{\value{NAT@total@cites}}
    {\if*\NAT@postnote*\else\NAT@cmt\NAT@postnote\global\def\NAT@postnote{}\fi}{}%
  \ifNAT@swa\else\if\relax\NAT@date\relax
  \else\NAT@@close\global\let\NAT@nm\@empty\fi\fi								
  \hyper@natlinkend}
\renewcommand\hyper@natlinkbreak[2]{#1}

\patchcmd{\NAT@citex}															
  {\ifNAT@swa\else\if*#2*\else\NAT@cmt#2\fi
   \if\relax\NAT@date\relax\else\NAT@@close\fi\fi}{}{}{}
\patchcmd{\NAT@citex}
  {\if\relax\NAT@date\relax\NAT@def@citea\else\NAT@def@citea@close\fi}
  {\if\relax\NAT@date\relax\NAT@def@citea\else\NAT@def@citea@space\fi}{}{}
\patchcmd{\NAT@cite}{\if*#3*}{\if*\NAT@postnote*}{}{}
\makeatother

\setlist[enumerate]{leftmargin=*,wide=0pt,itemsep=0pt,topsep=2pt}
\AtBeginDocument{
  \setlength\abovedisplayskip{4pt}
  \setlength\belowdisplayskip{4pt}
}
\begin{document}
\thispagestyle{empty}
\setcounter{page}{0}

\setcounter{footnote}{0}
\renewcommand{\thefootnote}{\fnsymbol{footnote}}
~\vspace*{-2cm}\\
\begin{center}\Large
    {\noindent
    Quantal Response Equilibrium with a Continuum of Types:\\
    Characterization and Nonparametric Identification
    }
\end{center}
\vspace*{1em}

\makebox[\textwidth][c]{
    \begin{minipage}{1.2\linewidth}
        \Large\centering
        Evan Friedman\footnotemark
        \quad \;\textcircled{r}\;\quad
        Duarte Gonçalves\footnotemark
    \end{minipage}
}
\setcounter{footnote}{1}\footnotetext{
    \setstretch{1} Department Economics, Paris School of Economics; \hyperlink{mailto:evan.friedman@psemail.eu}{\color{black}evan.friedman@psemail.eu}.
}
\setcounter{footnote}{2}\footnotetext{
    \setstretch{1} Department of Economics, University College London; \hyperlink{mailto:duarte.goncalves@ucl.ac.uk}{\color{black}duarte.goncalves@ucl.ac.uk}.
}

\blfootnote{
    We thank Olivier Compte and seminar participants at PSE, SAET 2023, ESA-SEOUL 2023, and the Virtual East Asia Behavioral Economics Seminar for helpful feedback.
    We are particularly grateful to the editor and two anonymous referees for their valuable comments and suggestions.
    \textcircled{r} denotes randomized authorship order \citep{Ray2018}.
    \\
	\emph{First posted draft}: 1 March 2023. 
    \emph{This draft}: 15 March 2024.
}

\setcounter{footnote}{0} \renewcommand{\thefootnote}{\arabic{footnote}}

\begin{center} \textbf{\large Abstract} \end{center}\vspace*{1em}
\noindent\makebox[\textwidth][c]{
    \begin{minipage}{.85\textwidth}
        \noindent
        Quantal response equilibrium (QRE), a statistical generalization of Nash equilibrium, is a standard benchmark in the analysis of experimental data.
        Despite its influence, nonparametric characterizations and tests of QRE are unavailable beyond the case of finite games.
        We address this gap by completely characterizing the set of QRE in a class of binary-action games with a continuum of types.
        Our characterization provides sharp predictions in settings such as global games, volunteer's dilemma, and the compromise game.
        Further, we leverage our results to develop nonparametric tests of QRE. 
        As an empirical application, we revisit the experimental data from Carrillo and Palfrey (2009) on the compromise game. 
        ~
        \\\\
        \textbf{Keywords:} quantal response; Bayesian games; global games; compromise game; nonparametric analysis.\\
        \textbf{JEL Classifications:} C44, C72, C92.
    \end{minipage}
}
\newpage

\section{Introduction}

Quantal Response Equilibrium (QRE) \citep{McKelvey1995}, a statistical generalization of Nash equilibrium (NE), has found significant success in explaining experimental data \citep*{Goeree2016}. 
In a QRE, players make probabilistic mistakes in best responding to their beliefs, but their beliefs are correct, taking into account the mistakes of others. 
This leads to systematic deviations from Nash equilibria that can capture a range of observed behavioral phenomena, such as turnout in large elections \citep{Levine2007} and overbidding in auctions \citep{Goeree2002b}.
However, while this solution concept has been influential, there are limited results on characterizing and testing QRE beyond the case of finite games.

In this paper, we consider a class of infinite games, those with binary actions and a continuum of types, for which we provide a complete characterization of the set of all quantal response equilibria (QRE). 
In this environment, a QRE is a function mapping types to the probability of taking a given action. 
Under a monotonicity condition on payoffs, our characterization is as follows: any QRE is a continuous, strictly monotone function such that uniform randomization implies indifference between actions. 
Further, we provide a converse: any such function is a QRE (for some underlying noise structure), and thus we fully describe the set of equilibria.
Using this result, we characterize QRE in a number of classic applications, including global games. 
Finally, we leverage our results to develop novel methods for nonparametric testing of QRE, which we apply to the experimental data of \citet{Carrillo2009} on the compromise game.

The games we study are symmetric, binary-action Bayesian games, with any number of players. 
Before taking an action, each player learns their \emph{type}, which potentially takes a continuum of values. 
This can either be a parameter of the utility function or a signal about some payoff-relevant state. 
To derive our main result, we assume only minimal structure on payoffs. 
Namely, we require that (1) (interim) expected payoffs are monotone in type whenever opponents' choice probabilities are also monotone in type, and that (2) for each of the two actions, there exists a type for which that action yields the higher payoff. 
While a simple class of infinite games, it is rich enough to include many games of significant theoretical interest.

The primitive of all QRE models is a \emph{quantal response function} --- the mapping from expected payoffs to a distribution over actions --- and a QRE obtains when all players' behavior is consistent with quantal response.
Following \citet{McKelvey1998}, we study \emph{agent QRE in symmetric strategies}, or simply \emph{QRE} for short. 
In such an equilibrium, each type represents an agent who acts independently, and different players with the same type have the same behavior. 
In this way, each type has only two actions, allowing us to focus on the role of an infinite type space (as opposed to an infinite strategy space). 
A QRE is thus a function mapping from types to the probability of taking a given action.

Rather than imposing parametric structure, in our approach, we take inspiration from recent work characterizing the set of all QRE based on minimal restrictions on the quantal response function.
\citet*{Goeree2005} define a \emph{regular QRE} as any for which the quantal response function satisfies \emph{monotonicity} and \emph{responsiveness}, as well as other technical axioms. 
These require that actions with higher expected payoffs are taken more often and that an all-else-equal increase in the payoff to some action means it is played even more often. 
Such restrictions are satisfied by a number of models generating stochastic choice. 
\citet{FriedmanMauersberger2022} study \emph{symmetric QRE}, a refinement of regular QRE, whereby the quantal response function also satisfies various symmetries across players and actions.
In this paper, we consider both regular and symmetric QRE.

Our main result fully characterizes the entire \emph{set} of QRE and enables the identification and recoverability of the QRE primitives from the data.
Specifically, we show that a function from types to choice probabilities is a QRE for some quantal response function satisfying the axioms if and only if the function is continuous, monotone, and uniform randomization implies indifference between actions.
Furthermore, symmetric QRE are exactly identified by functions satisfying these conditions together with an additional symmetry condition. 
This result is nonparametric in that it identifies the set of QRE as a subset in an infinite-dimensional space, without restricting the underlying quantal response function or assuming dependence on a finite-dimensional parameter.\footnote{
    For instance, while any function satisfying the conditions of our theorem is consistent with some quantal response function, it would typically not be consistent with a given parametric form such as logit.
}
We then leverage our characterization to develop a methodology to nonparametrically test if the data can be rationalized by QRE and estimate the quantal response function from the data.

Our result transforms the problem of finding a fixed point in an infinite-dimensional space to that of constructing a monotone function with a unique indifferent type who uniformly randomizes. 
While this may be a challenging problem, we show that it delivers simple and sharp predictions in the context of several classic applications. 
A key observation that makes the problem tractable is that expected payoffs often depend only on just a few features of the equilibrium strategy, for example its mean.\footnote{
    In our three applications, the relevant features are the mean, the distance of one's type to a particular reference type, and the quantile.
} 
In such cases, it is often easy to characterize the \emph{set of possible indifferent types} without fully specifying the underlying equilibrium, and then, for each indifferent type, construct the set of supporting mappings that satisfy the necessary features.

We then consider three applications: the volunteer's dilemma \citep{Diekmann1986} with a continuum of participation costs, global games \citep[e.g.][]{Carlsson1993,Morris1998} with a continuum signals about the state of the world, and the compromise game \citep{Carrillo2009} with a continuum of ``strengths.'' 
In each case, we apply our result to characterize the set of QRE, deriving a sharper characterization that depends on specific features of the games. 
The games were chosen to showcase a breadth of different arguments, with the goal of suggesting new applications.

We view our contribution as two-fold. 
Our first contribution is theoretical as our results expand the universe of games that are amenable to nonparametric QRE analysis. 
We also derive economic insights, showing in each of our applications the precise sense in which QRE deviates systematically from (Bayesian) Nash equilibrium. 
And while our focus is on characterizing sets of QRE nonparametrically, since the common parametric models are contained within the axiomatic families we study, our result implies bounds on these models as well.

Our second contribution is empirical. 
We show that our characterization results can be used as the basis for nonparametric tests of QRE.
Consider the common practice of fitting logit QRE to data. 
If the parametric model does \emph{not} fit well, it is unclear to what extent this is due to the logit structure or a general limitation of QRE.
By characterizing the set of QRE, our results allow us to nonparametrically test whether the data is consistent with \emph{some} QRE. 
This is tantamount to testing (1) if choice probabilities are monotone with respect to types, and (2) whether the type uniformly randomizing is indifferent between the two actions --- a simple moment condition.

As an empirical application, we revisit the experimental study of \citet{Carrillo2009} on the compromise game, which we re-analyze through the lens of our results. 
While we cannot reject monotonicity, we find a violation of the second condition: the type uniformly randomizing has a strictly higher expected payoff for one of the two actions.

This paper is organized as follows. 
In the remainder of this section, we discuss related literature. 
In \hyref{section:games}[Section], we introduce the family of games we consider, the definition of equilibrium, and provide general existence and characterization results. 
\hyref{section:applications}[Section] applies our results to characterize QRE in three applications: the
volunteer's dilemma, global games, and the compromise game. 
\hyref{section:empirics}[Section] introduces a methodology to test and estimate QRE in our class of games, and illustrates this by analyzing experimental data on the compromise game. \hyref{section:conclusion}[Section] concludes.

\subsection{Related literature}

In response to concerns that some forms of QRE might lack empirical content \citep[e.g.][]{Haile2008}, \citet*{Goeree2005} offer the first systematic consideration of more general, non-parametric forms. 
They introduce the axiomatic regular QRE and show it is falsifiable, which implies the same for the models it nests, namely logit QRE and, more generally, all ``structural'' QRE with i.i.d. errors.  
\citet{Goeree2018} introduced $M$ equilibrium for finite games,\footnote{
    See also the closely related rank-dependent choice equilibrium \citep{Goree2019_HB}.
} an explicitly set-valued concept that contains the set of all regular QRE.
Whereas QRE is defined for a given quantal response function and typically cannot be solved for in closed form, they show the set of all regular QRE can be characterized as a semi-algebra, i.e. in
terms of a finite number of polynomial inequalities.\footnote{
    Lemma 1 of \citet{Velez2023} establishes the same result for regular QRE using an alternative proof.
}  
This establishes the main insight that is relevant to our paper: by imposing only weak restrictions on quantal response, the resulting set of QRE is a tractable object. 
\citet{FriedmanMauersberger2022} refines regular QRE by augmenting
it with various forms of symmetry across players and actions, and
show how to analyze the resulting sets of equilibria.\footnote{
    \citet{friedman2021} provides some comparative static results for regular QRE augmented with \emph{translation invariance}.
} In binary-action games, the model is similarly tractable as regular QRE and implies much tighter bounds on the models nested within it, such as logit.

Whereas all of these previous papers focus on finite games, the main novelty of our paper is to provide characterization results for nonparametric QRE in a class of infinite games --- those with a continuum of types. 
This exercise is analogous to that undertaken for classes of finite games, but requires new methods altogether.

While fitting parametric QRE models and comparing their fit to other parametric models is common practice, there is surprisingly little work that develops formal tests. 
An exception is \cite{Melo2017}, which derives a test for structural QRE in sets of finite games.\footnote{
    See also \cite{Aguirregabiria2020} and \cite{Aguirregabiria2021} for approaches that require stronger distributional assumptions on errors.
} 
More recently, \cite{Hoelzemann2023} derive and test a necessary condition for QRE in sets of finite games under weaker conditions on the utility function.

Parametric QRE has been successfully applied to infinite games. 
In particular, logit QRE has been applied to specific games with continua of types that admit Bayesian Nash equilibria in threshold strategies. 
A prominent example is the experimental study of \citet{Levine2007}, who numerically approximate logit QRE for the incomplete information voting game of \citet{Palfrey1985}. 
Another example is the experimental study of \citet{Carrillo2009}, who numerically approximate logit QRE in the compromise game. 
Since this game is a special case of the class we study, we use their data in \hyref{section:empirics}[Section] to illustrate how to use our results in econometric testing for the adequacy of QRE to rationalize data.

In terms of games with continuous action spaces, \citet{Anderson2002} study logit QRE in a family of ``auctionlike'' games with ``payoff functions that depend on rank, such as whether a player's decision is higher or lower than another's.'' 
Here, a logit QRE is a choice density that satisfies a differential equation. 
While there is no closed-form expression for equilibrium strategies, \citet{Anderson2002} establish existence, uniqueness, and comparative statics. 
In a similar vein, \citet{Anderson2001} study logit QRE of a continuous minimum-effort coordination game, and \citet{Baye2001} study the parametric ``Luce'' QRE in a continuous Bertrand pricing game. 
We provide complementary results to these papers by characterizing the set of QRE in a non-overlapping class of games with a continuum of types.
A natural direction for future work is to extend our non-parametric analysis to games with larger action spaces.

QRE, which requires being able to assign a positive probability or density to all strategies, is not well-defined when the strategy space is a rich infinite-dimensional function space. 
Hence, we simplify the strategy space by considering interim or agent QRE \citep{McKelvey1998}. 
An alternative approach would be to impose a priori restrictions on strategies, as in \citet{Compte2019}. 
After imposing restrictions, possibly allowing for a family of stochastic strategies, \citet{Compte2019} study Nash equilibria of the restricted game. 
Alternatively, one could study QRE of the restricted game.\footnote{
    \citet{Carrillo2009}, in their logit QRE analysis of the compromise game, consider two versions. 
    The first is agent QRE, which is a parametric form of the model we study in this paper. 
    The second, which they refer to as ``cutpoint QRE'', imposes that each player only considers threshold strategies and then studies the QRE of this restricted game.
    \citet{Compte2019}, in several of their applications, restrict players to choose ``target'' actions that are implemented with exogenous trembles, but the resulting equilibria cannot be interpreted as QRE of a restricted game.
} In this way, it would be natural to combine non-parametric QRE methods with strategy restrictions, which could be a powerful approach for complex games with large strategy spaces.\footnote{
    See also \citet{Arad2019} for a theory of behavior in complex games.
}

\section{The games, equilibrium, and characterization}
\label{section:games}

We introduce the class of games, equilibrium concept, and restrictions on quantal response we consider. 
We then establish existence and characterize equilibria.

\subsection{Binary-action games with a continuum of types}

Let $I$ denote the set of players, which can be either finite (with at least 2 players) or a continuum. 
They play a symmetric binary-action game in which each player $i$ has the same binary-action set, $\mathcal Y:=\{0,1\}$.
There is an unknown (possibly degenerate) state of the world $\theta \in \Theta:= [0,1]$, distributed according to density $h$. 
Before taking an action, each player $i$ observes their private type $x_i\in \mathcal X:=[0,1]$, independently drawn conditional on the state according to density $f^\theta$; 
we require $f:=\int_{\Theta} f^\theta h(\theta)\diff \theta$ to have full support on $\mathcal X$.
Players act independently, conditional on the state of the world, and a player's payoff function is given by $u: \mathcal Y^I\times \mathcal X^I \times \Theta \to \mathbb R$, a measurable real-valued mapping depending on the players' action profile, the realized type profile, and the state of the world.

Anticipating the symmetric nature of the solution concept we consider, we focus on symmetric Lebesgue-measurable strategies, $\sigma: \mathcal X \to [0,1]$, where $\sigma(x_i)$ is the probability type $x_i$ takes action 1.
Given continuity properties imposed on the payoffs below and the fact that the distribution of types admits a density, we take the strategy space $\Sigma$ as the set of $L^1(\mathcal X)$ functions taking values in $[0,1]$, endowed with the $L^1$-norm $\|\cdot\|_{L^1}$.
We will denote the expected payoff to a player with type $x_i$ choosing action $y \in \mathcal Y$ given their opponents all follow strategy $\sigma$ as $\bar{u }^{y}_{x_i}(\sigma)$, formally given by
\[
    \bar{u }^{y}_{x_i}(\sigma):=\mathbb E_{\theta \sim h}[\mathbb E_{x_j \sim f^\theta, \forall j \ne i}[\mathbb E_{y_{j}\sim \sigma(x_j), \forall j \ne i}[u(y,y_{-i},x_i,x_{-i},\theta)]]\mid x_i].
\]

Further, define $\Delta \bar u_{x_i}(\sigma):=\bar{u}_{x_i}^1(\sigma)-\bar{u}_{x_i}^{0}(\sigma)$ as the corresponding expected utility difference between taking actions $1$ and $0$. 
Because of the symmetric nature of the environment, we henceforth omit player subscripts, using $x$ and $x'$ for types realized from an interim perspective. 

The above formulation is general enough to encompass many types of games. 
In particular, we note that a player's type $x$ can simply be a parameter of the utility function, or it can be a signal about the unknown state. 
We consider applications with both interpretations.

We impose the following restrictions on payoffs. Note that, in all cases, $\sigma$ refers to the symmetric strategy followed by all players:

\begin{enumerate}[label=(A\arabic*)]
    \item \textbf{Continuity}: 
    For all $y \in \mathcal Y$, $\bar{u}_{x}^{y}(\sigma)$ is jointly continuous in $(x,\sigma)$ with respect to the product topology. 
    \label{assumption:a1}
    
    \item \textbf{Payoff-responsiveness}: If $\sigma \in \Sigma$ is such that 
    $\sigma(x) > \sigma(x')$ for some $x< x'$, 
    then there exist 
    $\hat x \ne \hat x'$ satisfying 
    (i) $\sigma(\hat x) > \sigma(\hat x')$, and 
    (ii) $\bar{u}_{\hat x}^{1}(\sigma)\leq\bar{u}_{\hat x'}^{1}(\sigma)$ and 
    $\bar{u}_{\hat x}^{0}(\sigma)\geq\bar{u}_{\hat x'}^{0}(\sigma)$, with at least one of these inequalities strict. 
    \label{assumption:a2}

    \item \textbf{Payoff-monotonicity}: 
    If $\sigma \in \Sigma$ is such that 
    $\sigma$ is increasing, 
    then, for any $x<x'$, $\bar u_x^1(\sigma)\leq \bar u_{x'}^1(\sigma)$ and $\bar u_x^0(\sigma)\geq \bar u_{x'}^0(\sigma)$, with at least one of these inequalities strict.
    \label{assumption:a3}

    \item \textbf{Non-triviality}: 
    If $\sigma \in \Sigma$ is such that 
    $\sigma>1/2$ (resp. $\sigma<1/2$), 
    then there is $x\in \mathcal X$ such that $\Delta \bar u_{x}(\sigma)<0$ (resp. $\Delta \bar u_{x}(\sigma)>0$). 
    \label{assumption:a4}
\end{enumerate}

The first assumption \ref{assumption:a1} is continuity of expected payoffs, which is relatively innocuous, and guarantees that $\sigma$ will be continuous in equilibrium. 
The second assumption \ref{assumption:a2} imposes that, if $\sigma$ is decreasing at some point, then there exist two types such that one plays action 1 more often despite facing payoffs that are relatively less favorable to action 1. 
This will be shown to imply that $\sigma$ is strictly increasing in equilibrium, and is the weakest condition we have been able to formulate that guarantees this. 
In particular, we strove to allow for the existence of $\sigma$ such that expected payoffs are non-monotone in type, to be able to speak to many interesting applications.\footnote{
    An arguably natural strengthening of this condition that rules out such applications would be to have that
    for any $x< x'$ such that $\sigma(x) > \sigma(x')$, $\bar{u}_{x}^{1}(\sigma)\leq\bar{u}_{x'}^{1}(\sigma)$ and $\bar{u}_{x}^{0}(\sigma)\geq\bar{u}_{x'}^{0}(\sigma)$ with at least one of these inequalities strict.
} 
The third assumption \ref{assumption:a3} requires that monotone strategies imply payoff-monotonicity in type, which will allow us to characterize the full set of QRE. 
The last assumption \ref{assumption:a4} ensures some degree of strategic substitutability. 
Without this assumption, games admit equilibria in which there is an action that all types take more often than not. 
Such equilibria are not un-interesting, but QRE analysis turns out to be somewhat trivial. 
Hence, we think of this final assumption as a non-triviality constraint that allows us to focus on the most interesting cases.

We note that all Bayesian Nash equilibria in this class of games are in \emph{threshold strategies}.
We define Bayesian Nash equilibrium\footnote{
   We here abuse terminology by restricting attention to symmetric equilibria.
} 
as any strategy $\sigma\in \Sigma$ such that, for every type $x \in \mathcal X$, $\sigma(x)$ is a best response to $\sigma$, i.e. 
$\sigma(x) > 0 \Longrightarrow \Delta \bar u_x(\sigma)\geq 0$ and $\sigma(x) < 1 \Longrightarrow \Delta \bar u_x(\sigma)\leq 0$.
A threshold strategy is then defined as any strategy $\sigma\in \Sigma$ satisfying $\sigma=0$ on $x$<$x^*$ and $\sigma=1$ on $x>x^*$, for some threshold $x^*\in\mathcal X$.\footnote{
    Note that this definition allows for $\sigma(x)=0$ for all $x\in[0,1)$ or $\sigma(x)=1$ for all $x\in (0,1]$, which will be relevant for the compromise game of Section \hyref{subsection:compromise}[Section] in which the Bayesian Nash equilibrium prescribes $\sigma(x)=1$ for all $x\in(0,1]$.
}

\begin{proposition} 
    \label{proposition:BNE}
    For any game satisfying \ref{assumption:a1}-\ref{assumption:a4}, all Bayesian Nash equilibrium are in threshold strategies.
\end{proposition}
\begin{proof}
    See \hyref{subsection:proposition:BNE}[Appendix]. 
\end{proof}

Similar arguments have been used to prove existence of threshold equilibria in other settings, e.g. the incomplete information voting game with continuously distributed voting costs studied by \citet{Palfrey1985}. 
Hence, the class of games we study admit a familiar solution, and the QRE we characterize can be seen as stochastic generalizations of this solution.

\subsection{Quantal response equilibrium}

We assume that each type's behavior is governed by the same \emph{quantal response function} $Q:\mathbb R^{2}\to [0,1]$, which maps from expected payoffs to a mixed action that we identify with the probability of choosing action 1.\footnote{ 
    One can view $Q$ as the representative quantal response for a population of individuals with potentially heterogeneous quantal responses. When quantal responses arise from additive i.i.d. payoff disturbances and the action space is binary,  \citet{Golman2011} shows a representative quantal response emerges that is also based on additive i.i.d. payoff disturbances. 
}
We denote by $\bar u_x (\sigma):=(\bar u_x^1(\sigma),\bar u_x^0(\sigma))$ the vector of expected utilities for a player with type $x \in \mathcal X$ and strategy $\sigma:\mathcal X \to [0,1]$. 
Stated formally after imposing restrictions on $Q$, a quantal response equilibrium will be defined as a strategy $\sigma$ such that all types' behavior is consistent with quantal response: $\sigma(x)=Q(\bar u_x(\sigma))$ for all $x \in \mathcal X$.\footnote{
    Note that this corresponds to \emph{agent QRE} \citep{McKelvey1998}, which is common in the literature.
    Introduced to analyze extensive-form games, agent QRE treats the same player at different nodes --- or of different types --- as separate agents who mix independently. 
    This is often viewed as a simplification as each agent has a smaller strategy space than the player. 
    In this paper, each agent has exactly two actions. 
}

Without restrictions on $Q$, this poses almost no restrictions on observable behavior. 
Following existing literature, we impose weak restrictions or \emph{axioms} on $Q$. 
The axioms are defined for arbitrary finite numbers of actions, but we present them in a binary-action form. 
Throughout the paper, we always assume $Q$ satisfies the \emph{regularity} axioms \ref{assumption:r1}-\ref{assumption:r4} below, which are due to \citet{Goeree2005}. 

\begin{enumerate}[label=(R\arabic*)]
    \item \textbf{Interiority}: 
    $Q(v)\in(0,1)$ for all $v=(v^1,v^0)\in\mathbb R^{2}$. 
    \label{assumption:r1}
    \item \textbf{Continuity}: 
    $Q$ is continuous. 
    \label{assumption:r2}
    \item \textbf{Responsiveness}: 
    $\frac{\partial Q(v)}{\partial v^1}>0>\frac{\partial Q(v)}{\partial v^0}$ for all $v=(v^1,v^0)\in\mathbb R^2$.
    \label{assumption:r3}
    \item \textbf{Monotonicity}: 
    $v^1>v^0\iff Q(v)>1-Q(v)$.
    \label{assumption:r4}
\end{enumerate}

We now state the definition of the solution concept formally:
\begin{definition}
    Fix $Q$ satisfying \ref{assumption:r1}-\ref{assumption:r4}. 
    A \emph{quantal response equilibrium} (QRE) is a strategy $\sigma \in \Sigma$ such that 
    $\sigma(x)=Q(\bar u_x(\sigma))$ for all $x \in \mathcal X$.
\end{definition}

It is important to note that a QRE is a fixed point in a function space. 
Formally, define $\bar{u}(\sigma):=(\bar u_x (\sigma))_{x\in \mathcal X}$ as the vectors of expected payoffs faced by all types. 
Recalling that $\Sigma$ denotes the space of measurable functions mapping from $\mathcal X$ to $[0,1]$, we define the operator $q:\Sigma \to \Sigma$ to be such that, for all $x \in \mathcal X$, $q(\sigma)(x):=Q(\bar u_x(\sigma))$. 
An equivalent definition is then as a fixed point of $q$. That is, $\sigma \in \Sigma$ is a QRE if $\sigma = q(\sigma)$.

In addition to QRE, we consider a refinement that also imposes the \emph{symmetry} axioms \ref{assumption:s1}-\ref{assumption:s2} below, also introduced in \citet{Goeree2005}.

\begin{enumerate}[label=(S\arabic*)]
    \item \textbf{Translation invariance}: 
    $Q(v+\gamma e)=Q(v)$ for all $v=(v^{1},v^{0})\in\mathbb{R}^{2}$, $\gamma \in \mathbb R$ and $e=(1,1)$. 
    \label{assumption:s1}
    \item \textbf{Label independence}: 
    For any $v=(v^1,v^0),\tilde v=(\tilde v^1,\tilde v^0) \in \mathbb R^2$, if $v^1=\tilde v^0$ and $v^0=\tilde v^1$, then $Q(v)=1-Q(\tilde v)$.
    \label{assumption:s2}
\end{enumerate}

Following \citet{FriedmanMauersberger2022}, whenever $Q$ satisfies \ref{assumption:r1}-\ref{assumption:r4} and \ref{assumption:s1}-\ref{assumption:s2}, we refer to the resulting model as \emph{symmetric QRE} or \emph{sym-QRE}.

\begin{definition}
    Fix $Q$ satisfying  \ref{assumption:r1}-\ref{assumption:r4} and \ref{assumption:s1}-\ref{assumption:s2}. A \emph{symmetric quantal response equilibrium} (sym-QRE) is a strategy $\sigma: \mathcal X\to [0,1]$ such that $\sigma(x)=Q(\bar u_x(\sigma))$ for all $x \in \mathcal X$.
\end{definition}

The axioms \ref{assumption:r1}-\ref{assumption:r2} impose the key technical conditions --- that all actions are played with positive probability and that behavior is continuous in payoffs. 
\ref{assumption:r3}-\ref{assumption:r4} are the main behavioral axioms, imposing a weak form of rationality: higher payoff actions are played more often and an all-else equal increase in the payoff to some action leads to it being played even more often. 
\ref{assumption:s1} ensures that quantal response is invariant to adding the same constant to both payoffs, and \ref{assumption:s2} imposes that only actions' payoffs --- and not their labels --- matter for quantal response. 
\ref{assumption:s1}-\ref{assumption:s2} are not implied by \ref{assumption:r1}-\ref{assumption:r4}.
They do hold, however, under the common ``structural'' approach in which quantal response is induced by additive errors if the errors are exchangeable with respect to actions (weaker than i.i.d.) and invariant to the payoffs themselves. 
In virtually all parametrizations, \ref{assumption:r1}-\ref{assumption:r4} are satisfied; and in the large majority of applications, including the common logit QRE, \ref{assumption:s1}-\ref{assumption:s2} are also satisfied. 
In this paper, we study both QRE (\ref{assumption:r1}-\ref{assumption:r4}) and sym-QRE (\ref{assumption:r1}-\ref{assumption:r4} and \ref{assumption:s1}-\ref{assumption:s2}), which allows us to isolate the effects of symmetry.

\begin{remark} 
    While QRE is defined for a given $Q$, whenever we refer to some QRE or sym-QRE $\sigma$ without reference to any $Q$, it should be understood that there is some underlying $Q$ satisfying the relevant axioms. 
\end{remark}

\subsection{Existence and characterization}

Because of the infinite nature of the game, a QRE is a function $\sigma \in \Sigma = [0,1]^{\mathcal X}$ that is a fixed point of the operator $q$.
Our first step is to show that under general conditions, such a fixed point exists.
The crucial step of the proof is to invoke Schauder's fixed-point theorem, a generalization of Brouwer's fixed-point theorem for infinite dimensional spaces. 

\begin{lemma}
    \label{lemma:existence}
    Assume \ref{assumption:a1}-\ref{assumption:a3} and \ref{assumption:r2}-\ref{assumption:r4}.
    Then, the game admits a QRE, $\sigma=q(\sigma)$.
    Furthermore, any QRE is continuous and strictly increasing.
\end{lemma}
\begin{proof}
    See \hyref{subsection:lemma:existence}[Appendix]. 
\end{proof}

We next provide a characterization of QRE. 
We find that any QRE is continuous, strictly increasing, interior, and has a unique \emph{indifferent type} that uniformly randomizes.
Furthermore, these properties deliver a converse, and so we completely characterize the \emph{set of all} QRE. 

\begin{theorem}
    \label{theorem:rqre:characterization}
    Assume \ref{assumption:a1}-\ref{assumption:a4}.
    A strategy $\sigma \in \Sigma$ is a QRE    
    if and only if 
    (i) $\sigma$ is continuous and strictly increasing, (ii) $\sigma \in (0,1)$, and (iii) there exists a unique type $\tilde x \in(0,1)$ such that $\sigma(\tilde x)=\frac{1}{2}$ and $\Delta\bar u_{\tilde x}(\sigma)=0$. 
\end{theorem}
\begin{proof}
    \emph{Only if}: 
    That $\sigma$ is continuous and strictly increasing follows from \hyref{lemma:existence}[Lemma]. 
    Note that if payoffs satisfy \ref{assumption:a1}, since the relevant domain for $Q$ is a bounded set $\mathcal U \subset \mathbb R^2$ (defined formally in the proof of \hyref{lemma:existence}[Lemma]), from \ref{assumption:r1} it will also be the case that $\sigma(x)\in (0,1)$ for all $x$.
    Finally, we show that for any fixed point $\sigma=q(\sigma)$ there is a unique $\tilde x$ such that $\sigma(\tilde x)=1/2$.
    That there is at most one follows from the fact that $\sigma$ must be continuous and strictly increasing.
    Suppose now that there is no such type and instead $\sigma(x')>1/2$ $\forall x'\in \mathcal X$.
    Then by \ref{assumption:a4}, there exists $x$ such that $\Delta \bar u_x(\sigma)<0 \Longrightarrow \bar u_x^1(\sigma)<\bar u_x^0(\sigma)$. 
    From \ref{assumption:r4} we then get that $Q(\bar u_x(\sigma))=\sigma(x)<1/2$, a contradiction.
    A symmetric contradiction is obtained when assuming that $\sigma(x')<1/2$ $\forall x'\in \mathcal X$.
    Further, by \ref{assumption:r4}, $\sigma(\tilde x)=1/2\Longrightarrow \Delta\bar u_{\tilde x}(\sigma)=0$.
    
    \emph{If}: 
    From \ref{assumption:a3}, as $\sigma$ is strictly increasing, $\Delta\bar u_x(\sigma)$ is strictly increasing in $x$.
    Let $\delta: \mathcal X \to \mathbb R$ be given by $\delta(x):=\Delta\bar u_x(\sigma)$ and define $\tilde Q:[\delta(0),\delta(1)]\to [0,1]$ by $\tilde Q(d)=\sigma(\delta^{-1}(d))$, which is well-defined since $\delta$ is strictly increasing. 
    Extend this to the whole real line in any arbitrary way such that $\tilde Q:\mathbb R\to [0,1]$ is continuous, strictly increasing, and interior. 
    Finally, extend this to a quantal response function
$Q:\mathbb R^2\to (0,1)$ (i.e. defined over $\mathbb R^2$) by setting $Q(v^1,v^0)=\tilde Q(v^1-v^0)$.
    By construction, $Q$ satisfies \ref{assumption:r1}-\ref{assumption:r4} and $Q(\bar u_x(\sigma))=\tilde Q(\Delta\bar u_x(\sigma))=\sigma(x)\ 
   \forall x\in \mathcal X$.
\end{proof}

Intuitively, we find that a sym-QRE is a QRE with an additional symmetry condition across types. For ease of reference, we define \emph{symmetry} formally as a condition on $\sigma$ (and the expected payoffs induced by $\sigma$ and $u$). 
\begin{definition}
    A strategy $\sigma \in \Sigma$ is \emph{symmetric} if $\sigma(x)=1-\sigma(x') \Longleftrightarrow \Delta\bar u_x(\sigma) = - \Delta \bar u_{x'}(\sigma)$.
\end{definition}

The next result delivers a characterization of sym-QRE. 
It is the same as \hyref{theorem:rqre:characterization}[Theorem] but includes the above symmetry condition. 
\begin{theorem}
    \label{theorem:sym-qre:characterization}
    Assume \ref{assumption:a1}-\ref{assumption:a4}.
    A strategy $\sigma \in \Sigma$ is a sym-QRE if and only if 
    (i) $\sigma$ is continuous and strictly increasing, (ii) $\sigma \in (0,1)$, (iii) there exists a unique type $\tilde x \in(0,1)$ such that $\sigma(\tilde x)=\frac{1}{2}$ and $\Delta\bar u_{\tilde x}(\sigma)=0$, and (iv) $\sigma$ is symmetric. 
\end{theorem}
\begin{proof}
    \emph{Only if}: 
    For any QRE $\sigma=q(\sigma)$, properties (i)-(iii) follow from \hyref{theorem:rqre:characterization}[Theorem]; we now show (iv), the symmetry of $\sigma$.
    Let $\tilde Q(v^1-v^0):=Q((v^1-v^0,0))$.
    Note that, from \ref{assumption:s1}, $Q((v^1-v^0,0))=Q((v^1,v^0))$.
    Then, for any two payoff functions $u, u'$ such that $\Delta\bar u_x(\sigma)=\Delta \bar {u}'_x(\sigma)$ for all $\sigma$, we have $\tilde Q(\Delta \bar u_x(\sigma))=Q(\bar u_x(\sigma))=Q(\bar {u}'_x(\sigma))$ for all $\sigma$.
    \ref{assumption:s2} implies that $\tilde Q(\Delta \bar u_x(\sigma))=Q((\bar u^1_x(\sigma),\bar u^0_x(\sigma)))=1-Q((\bar u^0_x(\sigma),\bar u^1_x(\sigma)))=1-\tilde Q(-\Delta \bar u_x(\sigma))$, and so at any sym-QRE, $\sigma(x)=1-\sigma(x')\Longleftrightarrow \tilde Q(-\Delta \bar u_x(\sigma))=\tilde Q(\Delta \bar u_{x'}(\sigma)) \Longleftrightarrow -\Delta \bar u_x(\sigma)=\Delta \bar u_{x'}(\sigma)$, where the last equivalence follows from the fact that $\tilde Q$ is strictly increasing by \ref{assumption:r3}, proving $\sigma$ is symmetric.
    
    \emph{If}: Construct $Q:\mathbb R^2\to (0,1)$ where $Q(v^1,v^0)=\tilde Q(v^1-v^0)$ exactly as in the ``if'' direction of the proof of \hyref{theorem:rqre:characterization}[Theorem], except also we require that $\tilde Q (d) = 1- \tilde Q(-d)$, which is possible since $\sigma$ is symmetric.
    That $Q$ satisfies \ref{assumption:s1} and \ref{assumption:s2} (as well as \ref{assumption:r1}-\ref{assumption:r4}) and $Q(\bar u_x(\sigma))=\tilde Q(\Delta\bar u_x(\sigma))=\sigma(x)\, 
    \forall x \in \mathcal X$ follows from the construction.
\end{proof}

Any strategy satisfying the conditions of \hyref{theorem:rqre:characterization}[Theorems] or \ref{theorem:sym-qre:characterization} is, respectively, a QRE or sym-QRE for some $Q$; and any strategy violating these conditions is not, for any $Q$. 
In other words, these theorems characterize the \emph{set of} all QRE and sym-QRE, that is, taking the union over all $Q$ within the relevant class. 
A crucial point is that, for any given $Q$, a QRE is defined as a fixed point in a high-dimensional space that typically has no closed-form solution. 
However, by taking the union over all $Q$, the set of QRE admits a tractable characterization that does not involve any fixed points. 
The same insight was previously shown by \citet{Goeree2018} and \citet{Goree2019_HB} in the context of finite games. 
Our results show that this insight is not limited to finite games, and therefore potentially much more general. 
We emphasize, however, that our characterization substantially differs from that in finite games, where the set of QRE can be characterized as a semi-algebra, i.e. in terms of finite polynomial inequalities.

As we discuss in \hyref{section:empirics}[Section], \hyref{theorem:rqre:characterization}[Theorems] and \ref{theorem:sym-qre:characterization}, by giving simple necessary and sufficient conditions, also pave the way for a general methodology to nonparametrically test the ability of QRE and sym-QRE to rationalize data.

In specific applications, we derive even sharper characterizations of the set of QRE by constructing strategies satisfying the conditions of \hyref{theorem:rqre:characterization}[Theorems] and \ref{theorem:sym-qre:characterization}. 
While this does not involve fixed point calculations, such constructions may not be entirely straightforward: one must construct monotonic $\sigma$ such that there is a unique indifferent type $\tilde x$ who uniformly mixes. 
However, as we show in \hyref{section:applications}[Section], this problem is further simplified by the fact that, in applications, expected payoffs often do not depend on $\sigma$ in its entirety, but rather on just a few of its properties.
For example, it could be that the payoffs to type $x'$ depend on $\sigma$ only through a specific statistic, such as its mean or a particular quantile. 
In such cases, the problem becomes particularly tractable as one may be able to characterize the \emph{set of indifferent types} without being precise about the supporting strategies, and only then construct the set of strategies (satisfying the relevant conditions) that can support each indifferent type.

\begin{remark}
    It is immediate that, in light of \ref{assumption:a1} and \ref{assumption:r1}-\ref{assumption:r4}, one can relax \ref{assumption:a2}-\ref{assumption:a4} and expand the class of games \hyref{theorem:rqre:characterization}[Theorems] and \ref{theorem:sym-qre:characterization} apply to.
    For instance, relaxing \ref{assumption:a2}-\ref{assumption:a4} by restricting attention to strategies $\sigma$ that are continuous and everywhere map to interior probabilities leaves the results unchanged.
\end{remark}

\subsection{Identification}

Our last general result pertains to the identification of the quantal response function $Q:\mathbb R^2\to [0,1]$. While our characterization of the set of QRE makes no reference to the underlying quantal response functions, we show how to partially recover $Q$ from equilibrium play.

To this end, for any QRE $\sigma$, define 
$V(\sigma):=\{(v^1,v^0) \in \mathbb R^2\mid \exists x \in \mathcal X:  \Delta \bar u_x(\sigma) = v^1-v^0\}$ to be the set of expected payoff vectors that arise in equilibrium as well as all translations of such vectors. For all $v \in V(\sigma)$, we also define $x(v) \in \mathcal X$ as the unique type satisfying $\Delta\bar u_{x(v)}(\sigma) = v^1-v^0$. 

For any QRE $\sigma$, our next result provides a construction for $Q\vert_{V(\sigma)}$ that is consistent with $\sigma$. 
If $\sigma$ is also a sym-QRE, the same construction provides the \emph{unique} such $Q\vert_{V(\sigma)}$.

\begin{theorem}
    \label{theorem:recover}
    Assume \ref{assumption:a1}-\ref{assumption:a4}.
    \begin{enumerate}[label=(\arabic*)]
        \item 
        If $\sigma$ is a QRE, then there is a $Q: \mathbb R^2 \to [0,1]$ (satisfying \ref{assumption:r1}-\ref{assumption:r4} and \ref{assumption:s1}) with
        $Q\vert_{V(\sigma)}(v)=\sigma(x(v)) \ \forall v \in V(\sigma)$ such that $\sigma(x)= Q(\bar u_x(\sigma))\,\ \forall x \in \mathcal X$.
     
        \item 
         If $\sigma$ is a sym-QRE, then all $Q: \mathbb R^2 \to [0,1]$ such that $\sigma(x)= Q(\bar u_x(\sigma))\,\ \forall x \in \mathcal X$ satisfy $Q\vert_{V(\sigma)}(v)=\sigma(x(v))\ \forall v \in V(\sigma)$.
    \end{enumerate}
\end{theorem}
\begin{proof}
    \emph{(1)}: This follows exactly from the construction in the ``if'' direction of the proof of \hyref{theorem:rqre:characterization}[Theorem]. 
    \emph{(2)}: This is the construction in the ``if'' direction of the proof of \hyref{theorem:sym-qre:characterization}[Theorem]. 
    That this is unique follows from the fact that $\sigma$ uniquely identifies $Q$ restricted to the payoff vectors observed in equilibrium, which, by \ref{assumption:s1}, extends uniquely to $Q\vert_{V(\sigma)}$.
\end{proof}

Hence, we may recover a significant portion of $Q$ from equilibrium play. 
The key behind this result is the game's payoff richness: in any QRE, we observe the expected payoffs and associated mixed actions for the entire continuum of types. 
Hence, unlike for finite games, the underlying $Q$ is significantly pinned down by observable behavior. 
With the auxilliary assumption of translation invariance, as in sym-QRE, $Q$ is completely pinned down over all of $V(\sigma)$, a set of positive measure.

\section{Applications}
\label{section:applications}

We consider three games: the volunteer's dilemma, the compromise game,
and a global game. 
The games were chosen to showcase a broad range of possible applications.  
While we invoke \hyref{theorem:rqre:characterization}[Theorems] and \ref{theorem:sym-qre:characterization} in all applications, the arguments are unique in each case.\footnote{
    In \ref{section:conditions}, we show that the payoffs satisfy the regularity needed to obtain the results in \hyref{theorem:rqre:characterization}[Theorems] and \ref{theorem:sym-qre:characterization}, as one can further relax \ref{assumption:a1}-\ref{assumption:a4}.
    For example, in the global games application, rather than establishing \ref{assumption:a2}, we show directly that all QRE must be strictly increasing. 
    Since \ref{assumption:a2} is only used to establish this, existence and characterization results go through unchanged. 
    It is also the case that for this, and all other applications we consider, the unique symmetric Bayesian Nash equilibrium is in threshold strategies. 
}

\subsection{Volunteer's dilemma}
\label{subsection:volunteers}

A huge literature studies voluntary contribution toward a public good when contributions are costly and there is an incentive to ``free ride.'' 
The volunteer's dilemma is a simple variant of this game in which the public good is provided if and only if at least one player contributes, so that it is optimal to contribute if and only if no other player does.\footnote{
    The volunteer's dilemma is a special case of the threshold public goods game of \citet{Palfrey1984}, which is analyzed using QRE by \citet{GH_2005}.
} The game was introduced by \citet{Diekmann1986}, and has since been revisited in many theoretical and experimental studies. 
We study a novel variant in which the contribution cost is continuously-distributed private information.

Two players simultaneously decide whether to \emph{volunteer} to perform a task (action 0) or to \emph{abstain} (action 1). 
If at least one player volunteers, both receive $B\in(1,2)$. 
However, volunteering is costly. 
A player's private cost $x$ is uniformly distributed on $\mathcal X=[0,1]$, i.i.d. across players. 
Let $\sigma(x)$ denote the probability that a player with cost $x$ abstains.\footnote{
    In the version introduced by \citet{Diekmann1986}, there are $N$ players, each of whom has the same cost.
}

Volunteering ensures that the benefit is received and the cost is paid, so the value for type $x$ of volunteering is $B-x$. 
By abstaining, type $x$ forgoes the cost, but only benefits if the other player volunteers; hence, this yields an expected payoff of $B\int_0^1 1-\sigma(x')\diff x'=B\,\mathbb E[1-\sigma(x')]$.
The difference in payoffs between abstaining and volunteering is:
\begin{align*}
    \Delta \bar u_{x}(\sigma) & =x-B \mathbb E[\sigma(x')].
\end{align*}
Hence, the payoff difference depends on $\sigma$ only through the
mean $\mathbb E[\sigma(x')]= \int_{0}^{1}\sigma(x')\diff x'$ and is additively separable in type $x$. 
These features make the analysis
particularly simple.

As a benchmark, consider first the (essentially) unique (symmetric) Bayesian Nash equilibrium, which is in threshold strategies: $\sigma^{BNE}(x)={1}\left\{x>\frac{B}{B+1}\right\}$.\footnote{
    The type $x=\frac{B}{B+1}$ is indifferent and may mix arbitrarily.
} Low-cost types volunteer, high-cost types abstain, and there is a unique indifferent type $\tilde x^{BNE}=\frac{B}{B+1}$ that can mix arbitrarily. 

Intuitively, by injecting noise as in QRE, this step function will be smoothed out, and the flexibility in the admissible noise structures leads to a range of possible indifferent types.

Let $\tilde x(\sigma)$ denote the type such that $\Delta \bar u_{\tilde x(\sigma)}(\sigma)=0$, i.e. $\tilde x(\sigma):=\inf\{x\in (0,1)\mid \Delta \bar u_x(\sigma)>0\}$.
Let $\tilde X$ denote the set of indifferent types for QRE, i.e. $\tilde X:=\{\tilde x(\sigma) \mid \sigma \in \Sigma \text{ is a QRE }\}$.
We then obtain the following characterization:
\begin{proposition}
    \label{proposition:rqre:volunteer}
    A strategy $\sigma \in \Sigma$ is a QRE    
    if and only if 
    (i) $\sigma$ is continuous and strictly increasing, (ii) $\sigma \in (0,1)$, and (iii) there exists a unique indifferent type $\tilde x \in \tilde X=\left(\frac{B}{B+2},2\frac{B}{B+2}\right)$ such that $\sigma(\tilde x)=\frac{1}{2}$ and $\tilde x=B \mathbb E[\sigma(x')]=B\int_{0}^{1}\sigma(x')\diff x'$. 
\end{proposition}
\begin{proof}
    See \hyref{subsection:proposition:rqre:volunteer}[Appendix]. 
\end{proof}

QRE is very flexible relative to Bayesian Nash equilibrium. 
Much of this flexibility comes from the fact that, while types must tend to take the action that yields the higher payoff (and uniformly mix when indifferent), they may still be biased in favor of a particular action. 
For example, if $\sigma(x)<1-\sigma(x')$ for some $x<\tilde x<x'$ and $\vert\Delta\bar u_{x}(\sigma)\vert \leq \vert\Delta \bar u_{x'}(\sigma)\vert$, then there is a (local) bias in favor of volunteering. 
Sym-QRE, by imposing symmetry, rules out precisely these biases.
In the sequel, we define $\tilde S:=\{\tilde x(\sigma) \mid    \sigma \in \Sigma \text{ is a sym-QRE }\}$ as the set of indifferent types for sym-QRE.

\begin{proposition}
    \label{proposition:sym-qre:volunteer}
    A strategy $\sigma \in \Sigma$ is a sym-QRE    
    if and only if 
    (i) $\sigma$ is continuous and strictly increasing, (ii) $\sigma \in (0,1)$, (iii) there exists a unique indifferent type $\tilde x \in \tilde S = \left(\frac{B}{B+1},\frac{B}{2}\right)$ such that $\sigma(\tilde x)=\frac{1}{2}$ and $\tilde x=B \mathbb E[\sigma(x')]=B\int_{0}^{1}\sigma(x')\diff x'$, and (iv) $\sigma$ is symmetric, satisfying $\sigma(\tilde x + \delta)=1-\sigma(\tilde x-\delta)$ for any $\delta \in[0,1-\tilde x]$. 
\end{proposition}
\begin{proof}
    See \hyref{subsection:proposition:sym-qre:volunteer}[Appendix]. 
\end{proof}

\begin{figure}[!ht]
    \centering
    \makebox[1.2\linewidth][c]{
    \begin{minipage}{1.1\linewidth}\centering
        \hspace*{-7em}\begin{subfigure}{.42\linewidth}
            
            \includegraphics[width=1\linewidth]{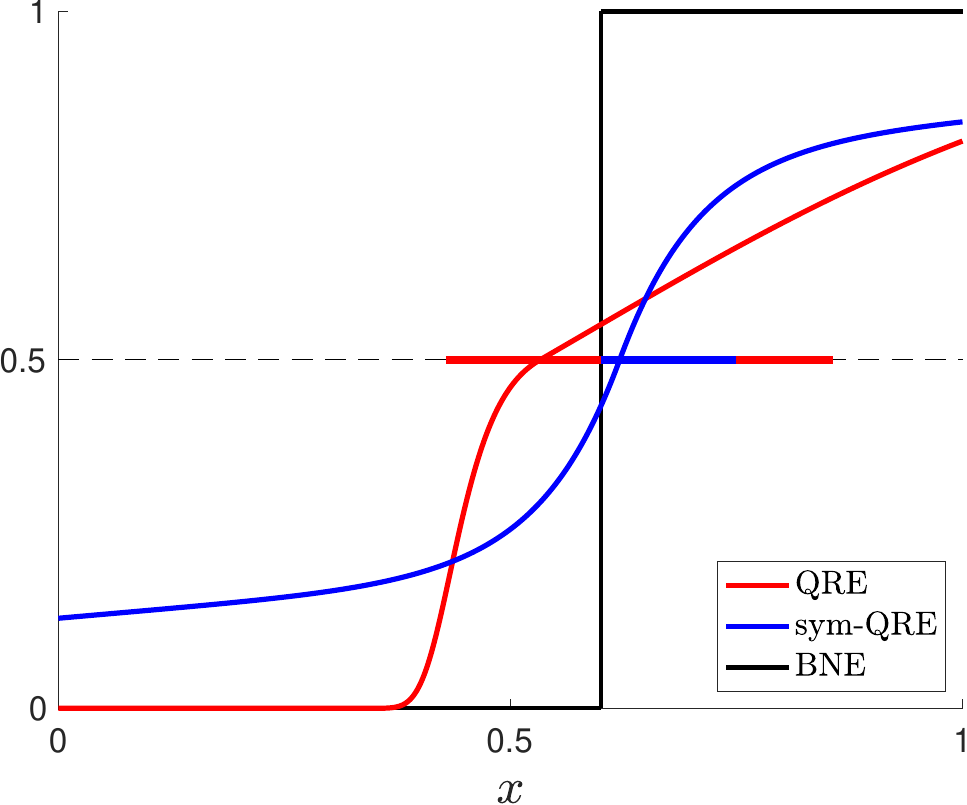}
            \caption{Equilibria}
            \label{figure:volunteer-QRE}
        \end{subfigure}
        \begin{subfigure}{.42\linewidth}
            
            \hspace{.15in}
            \includegraphics[width=1.0\linewidth]{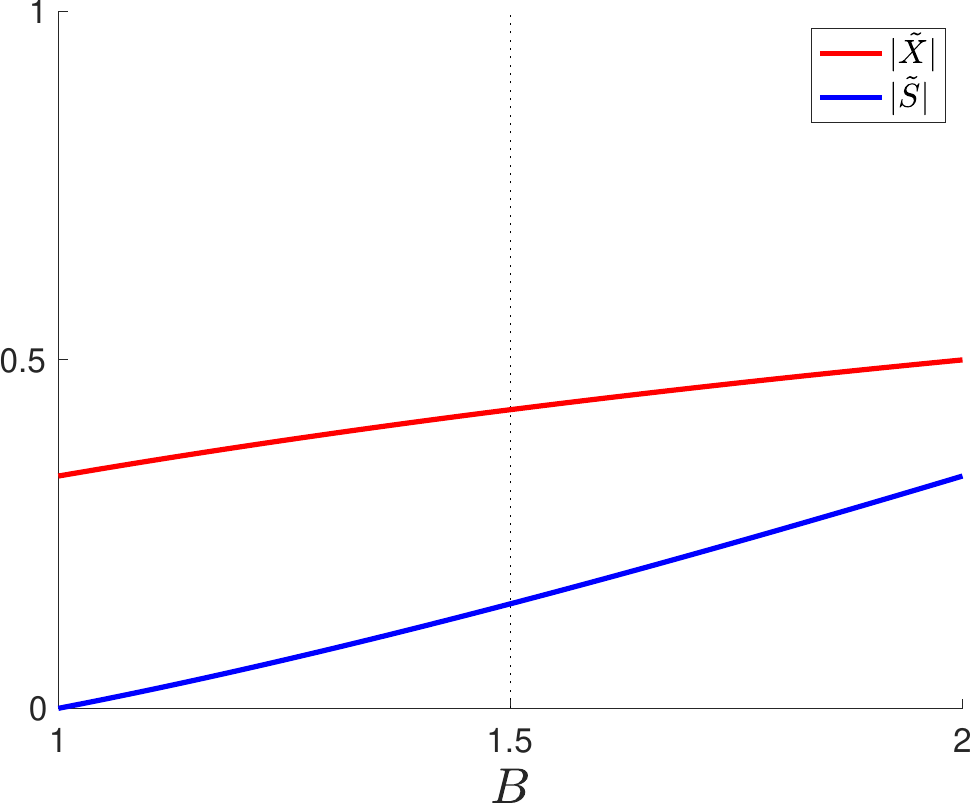}
            \caption{Measure of Indifferent Types}
            \label{figure:volunteer-types}
        \end{subfigure}
    \end{minipage}
    }
    \begin{minipage}{1\linewidth}
        \small
        \vspace*{.5em}
        \caption{QRE and sym-QRE in the Volunteer's dilemma}
        \label{figure:volunteer}
        \vspace*{-1.0em}
        \singlespacing \emph{Notes}: 
        Panel \ref{figure:volunteer-QRE} depicts the Bayesian Nash equilibrium (black) as well as illustrative QRE (red) and sym-QRE (blue), for $B=1.5$. The thick horizontal line at one-half indicates the set of indifferent types under QRE; the part in blue gives the set of indifference types under sym-QRE.
        Panel \ref{figure:volunteer-types} exhibits the measures of the sets of possible indifferent types for QRE (red) and sym-QRE (blue) for $B \in (0,2)$.
    \end{minipage}
\end{figure}

The left panel of \hyref{figure:volunteer}[Figure] shows an illustrative QRE (in red) and a sym-QRE (in blue) for the case that $B=1.5$. 
The thick horizontal lines at $\frac{1}{2}$ represent the sets of possible indifferent types $\tilde{X}=(\frac{B}{B+2},\frac{2B}{B+2})$ and $\tilde{S}=(\frac{B}{B+1},\frac{1}{2}B)$. While all sym-QRE satisfy symmetry, this particular QRE is drawn with a bias in favor of volunteering: there exists $x<\tilde x<x'$ such that $\sigma(x)<1-\sigma(x')$ and $\vert\Delta \bar u_{x}(\sigma)\vert\leq\vert\Delta \bar u_{x'}(\sigma)\vert$.

We find that, while the flexibility in both QRE models gives rise to a range of possible behaviors, sym-QRE gives much more precise predictions.
For instance, consider the measures of $\tilde{X}$ and $\tilde{S}$, which are $\vert\tilde{X}\vert=\frac{B}{B+2}$ and $\vert\tilde{S}\vert=\frac{B(B-1)}{2B+2}$, respectively. 
Plotting these measures in the right panel of \hyref{figure:volunteer}[Figure] as a function of $B$, we see that they are larger under QRE than sym-QRE for all values of $B$. 
There are also important qualitative differences between QRE and sym-QRE. 
We find that $\tilde x^{BNE}=\frac{B}{B+1}$, the indifferent type under Bayesian Nash equilibrium, is always in the interior of $\tilde{X}$, whereas $\tilde x^{BNE}$ is precisely the infimum of $\tilde{S}$. 
This gives a sense in which sym-QRE leads to more systematic deviations from Nash equilibrium.

In the volunteer's dilemma, as a direct consequence of the indifference condition, there is a one-to-one mapping between the indifferent type $\tilde x$ and the ex-ante probability of abstention: $\mathbb E[\sigma(x')]=\tilde x/B$. The characterization results can therefore be reframed in terms of this easier-to-interpret equilibrium object, leading to the following result.  
Specifically, the set of attainable values of $\mathbb E[\sigma(x')]$ is $(1/(B+2),2/(B+2))$ for QRE and $(1/(B+1),1/2)$ for sym-QRE.
Hence, sym-QRE always yields a lower probability of abstaining than in Bayesian Nash equilibrium, whereas QRE can yield a lower or higher value depending on the direction of bias.

\textbf{Initial discussion.} 
With the volunteer's dilemma as an example, we discuss four general points. 

First, as already mentioned, this is one of the simplest possible applications because payoffs depend on $\sigma$ only through its expectation $\mathbb E [\sigma]$ and payoffs are additively separable in type $x$. We show in subsequent sections, however, that we may still obtain precise characterization results without these features. 

Second, we show in \ref{section:volunteer:logit} that, in this example, the set of indifferent types attainable in sym-QRE can also be attained by logit QRE by varying the precision parameter $\lambda$. 
We conjecture that this is true in all of our applications and under fairly weak conditions. 
However, for any given indifferent type, there will be many strictly increasing, symmetric strategies $\sigma$ that make that type indifferent and uniformly randomizing, as required by our characterization, but almost all such strategies would not be consistent with logit QRE. 
Moreover, the logit QRE structure itself is unhelpful in analytically deriving the set of indifferent types.

Third, one may expect that the set of indifferent types will always be an interval with one endpoint being the threshold type under Bayesian Nash equilibrium (i.e. when there is very little noise) and the other endpoint being the type that is indifferent when all types uniformly randomize (i.e. when there is a lot of noise). 
In all of our applications, we find that this is, in fact, precisely the set of indifferent types under sym-QRE, and we conjecture this would be the case more generally. 
With regular QRE, however, we find that the set of indifferent types is much larger in the volunteer's dilemma. 
This is because QRE allows for the types above and the types below the indifferent type to have very different noise levels. 
Interestingly, while the set of equilibrium strategies is always larger under QRE, this additional flexibility does not always translate into a larger set of indifferent types, as we show in one of the later applications.

Finally, because the space of all strategies is an (infinite-dimensional) function space, and the set of QRE is a subset of that space, it is unclear how to quantify the ``size'' of the set of QRE in our setting. 
This contrasts with the case of finite games where the (Lebesgue) measure of the set of QRE can be directly computed (see \citet{Goeree2018}). 
However, there is a sense in which QRE excludes ``most'' strategies. 
In general, every QRE (sym-QRE) must be continuous and strictly increasing and satisfy $\sigma^{-1}(1/2)\in\tilde{X}$ $(\tilde{S})$; and the measure of these sets can be very small.\footnote{
    As shown in \hyref{figure:volunteer}[Figure], the measure of $\tilde{S}$ in the Volunteer's dilemma can be made arbitrarily close to 0 by taking $B\rightarrow 1^+$.
} There are additional restrictions that are specific to each game. 
In the volunteer's dilemma, for example, any given indifferent type pins down the average action of all supporting strategies. 
Hence, if we identify all strategies with their expectation, we can say that almost all strictly increasing strategies satisfying $\sigma^{-1}(1/2)\in \tilde X$ $(\tilde S)$ are \emph{not} consistent with QRE (sym-QRE).

\subsection{Global games}
\label{subsection:global}

Global games, first studied by \citet{Carlsson1993}, offer a tractable model of many complex economic problems. 
Examples include currency attacks and bank runs, for which a player's payoffs depend on their own action, the actions of others, and some economic fundamental, summarised by state $\theta$. 
The distinguishing feature of global games is that, when $\theta$ is observable, there is a multiplicity of equilibria that is eliminated when, instead, each player only observes a private signal of $\theta$ (no matter how precise).

The particular variant of global games we study is most similar to that of \citet{Morris1998}.\footnote{
    In \citet{Morris1998}, $\theta$ represents the strength of the regime --- the required mass of attacking players for regime change. 
    In our version, it represents the direct value of attacking, regardless of whether or not the attack succeeds. 
} A continuum of players decide whether to \emph{attack} a regime (e.g. a currency peg) (action 1) or to \emph{abstain} (action 0). 
The attack is successful if and only if (strictly) more than $1/2$ of the mass of players attack. 
Attacking always yields an uncertain payoff $\theta$, uniformly distributed on $\Theta=[0,1]$, but, if it fails, attackers pay an additional penalty of $c\in (0,1)$.
Abstaining yields the safe payoff of $k>0$. 
Note that, were $\theta$ known, attacking is strictly dominated if $\theta<\underline \theta:=k$ and it is strictly dominant if $\theta>\overline \theta:=k+c$.
To avoid trivial cases, we assume $\underline \theta,\overline \theta \in (0,1)$.

Prior to taking an action, each player privately observes their type $x\in \mathcal X = [0,1]$, which is a signal about $\theta$, identically and independently distributed across players conditional on $\theta$.
We assume that, given $\theta$, $x$ is distributed according to density $f(\cdot|\theta;\epsilon)$, parametrized by  $\epsilon$; for ease of notation we will omit the dependence on $\epsilon$.
For simplicity we consider the case in which $x$ is uniformly distributed and corresponds to the posterior mean about $\theta$, that is, $x=\mathbb E[\theta |x]$.
Specifically, we assume that for $\theta \in [\epsilon,1-\epsilon]$, $x$ is uniformly distributed on $[\theta-\epsilon,\theta+\epsilon]$, whereas if $\theta<\epsilon$ or $\theta>1-\epsilon$, we have that $x\sim U[0,2\theta]$ and $x\sim U[2\theta-1,1]$, respectively.
Parameter $\epsilon>0$  captures the imprecision of the signal, which we require to be ``small,'' satisfying $\epsilon<\min\{\underline \theta,1-\overline \theta\}$.

The goal of this section will be to use our results to provide a sharp characterization of the entire set of QRE, which are continuous and strictly increasing.\footnote{
    Rather than showing \ref{assumption:a1} and \ref{assumption:a2} do hold, in \hyref{subsection:proposition:unique:global}[Appendix] we prove directly and independently that all QRE must be continuous and strictly increasing; and therefore \hyref{theorem:rqre:characterization}[Theorems], \ref{theorem:sym-qre:characterization}, and \ref{theorem:recover} hold unchanged.
}
Differently from the previous application, here a player's private type corresponds to their private signal about the state.
We denote by $\sigma(x)$ the probability that a player with signal $x$ chooses to attack.

We recall that there will be an essentially unique (symmetric) Bayesian Nash equilibrium, given by $\sigma^{BNE}(x)=1\{x>(\overline \theta + \underline \theta)/2\}$.\footnote{
    The type $x=\frac{\underline \theta + \overline \theta}{2}$ is indifferent and may mix arbitrarily.
} 
In order to characterize the set of QRE, let us first introduce some definitions that will be used in our analysis.

Let $\overline \sigma(\theta)$ denote the mass of players attacking given strategy $\sigma$ and state $\theta$, i.e. 
$\overline \sigma(\theta):=\mathbb E[\sigma(x')|\theta]=\int_0^{1}\sigma(x')f(x'|\theta)\diff x'$.
While an attack fails at $\theta$ if and only if $\overline \sigma(\theta)\leq 1/2$, players do not observe $\theta$ but a signal $x$ about $\theta$.
Let $P(x,\sigma)$ denote the \emph{subjective failure probability} --- the belief that a player with type $x$ holds about the attack's success given strategy $\sigma$. This is defined as 
$P(x,\sigma):=\mathbb E[1\{\overline \sigma(\theta)\leq 1/2\}|x]=\frac{1}{2\epsilon}\int_{x-\epsilon}^{x+\epsilon}1\{\overline \sigma(\theta)\leq 1/2\}\diff \theta$.

For any strictly increasing $\sigma$ --- and, therefore, for any QRE --- there will be a \emph{threshold state} $\theta^*(\sigma)$ such that $\overline \sigma(\theta^*(\sigma))= \frac{1}{2}$ and so attacks will fail whenever $\theta \leq \theta^*(\sigma)$ and succeed when $\theta > \theta^*(\sigma)$.
Hence, for any such $\sigma$, $P(x,\sigma)=1$ if $x\leq \theta^*(\sigma)-\epsilon$, $P(x,\sigma)=0$ if $x\geq \theta^*(\sigma)+\epsilon$, and $P(x,\sigma)=\frac{\theta^*(\sigma)-x+\epsilon}{2\epsilon}$ for intermediate types $x\in (\theta^*(\sigma)-\epsilon,\theta^*(\sigma)+\epsilon)$.
Moreover, for any $\sigma$, we can easily express the expected payoff difference as 
\begin{align*}
    \Delta \bar u_x(\sigma)=\mathbb E[\theta|x]-c P(x,\sigma)-k=x-\underline \theta-(\overline \theta-\underline \theta)P(x,\sigma).
\end{align*}
As such, in any QRE, any player's expected payoff difference depends only on their type $x$ and the threshold $\theta^*(\sigma)$.
An implication is that any given indifferent type $\tilde x$ (recall $\tilde x$ satisfies $\Delta \bar u_{\tilde x}(\sigma)=0$) pins down both $P(\tilde x,\sigma)$ and $\theta^*(\sigma)$, and therefore the payoffs for all types $x\in \mathcal X$.
It is this limited dependence of payoffs on $\sigma$ that makes the problem tractable. 
In particular, we make use of the following lemma, which follows immediately from the expressions for $\Delta \bar u_{\tilde x}(\sigma)=0$ and $P(\tilde x,\sigma)$.
\begin{lemma}
    \label{lemma:global:failure}
    For any continuous and strictly increasing QRE $\sigma$, 
    (1) $\tilde x \in (\underline \theta,\overline \theta)$,
    (2) $P(\tilde x, \sigma)=\frac{\tilde x- \underline \theta}{\overline \theta-\underline \theta}$,
    (3) $\theta^*(\sigma)=\tilde x\left(\frac{2\epsilon}{\overline \theta-\underline \theta}+1\right)+\epsilon\left(\frac{\overline \theta+\underline \theta}{\overline \theta-\underline \theta}\right) \in (\tilde x-\epsilon,\tilde x+\epsilon)$, and 
    (4) for $x<x'$, $\Delta \bar u_{x'}(\sigma)-\Delta \bar u_x(\sigma)=x'-x + \frac{(\overline \theta-\underline \theta)}{2\epsilon}|M|>0$, where $M=[\theta^*(\sigma)-\epsilon,\theta^*(\sigma)+\epsilon]\cap [x,x']$.
\end{lemma}
\begin{proof}
    See \hyref{subsection:lemma:global:failure}[Appendix]. 
\end{proof}

Our first result establishes uniqueness. 
The intuition is similar to that of the classic result of \citet{Morris1998}: without a publically observed $\theta$, perfect coordination is impossible. 
\begin{proposition}
    \label{proposition:unique:global}
    For any $Q$ satisfying \ref{assumption:r1}-\ref{assumption:r4}, there is a unique QRE, which is continuous and strictly increasing. 
\end{proposition}
\begin{proof}
    See \hyref{subsection:proposition:unique:global}[Appendix]. 
\end{proof}
With uniqueness established for every quantal response function, we now provide a sharp characterization of the set of QRE. 
\begin{proposition}
    \label{proposition:rqre:global}
    A strategy $\sigma \in \Sigma$ is a QRE 
    if and only if 
    (i) $\sigma$ is continuous and strictly increasing, (ii) $\sigma \in (0,1)$, and (iii) there exists a unique indifferent type $\tilde x \in \tilde X=(\underline \theta,\overline \theta)$ such that $\sigma(\tilde x)=\frac{1}{2}$ and $\tilde x=P(\tilde x, \sigma)(\overline \theta-\underline \theta)+\underline \theta$. 
\end{proposition}
\begin{proof}
    See \hyref{subsection:proposition:rqre:global}[Appendix]. 
\end{proof}

For QRE, the set of possible indifferent types is very large; it is the entire set of types for which neither action is dominant. 
Furthermore, since the threshold state $\theta^*(\sigma)$ depends only on $\sigma(x')$ for $x'\in(\theta^*(\sigma)-\epsilon,\theta^*(\sigma)+\epsilon)$, and the expected payoffs to each type $x$ depend only on its distance to $\theta^*(\sigma)$, there are many $\sigma$ that are consistent with any given indifferent type. 
By contrast, all sym-QRE are associated with a unique indifferent type corresponding to the indifferent type in the unique Bayesian Nash equilibrium, and $\sigma$ must be symmetric about that indifferent type.

\begin{proposition}
    \label{proposition:sym-qre:global}
    A strategy $\sigma \in \Sigma$ is a sym-QRE 
    if and only if 
    (i) $\sigma$ is continuous and strictly increasing, (ii) $\sigma \in (0,1)$, and (iii) there exists a unique indifferent type $\tilde x \in \tilde S=\left\{\frac{\overline \theta + \underline \theta}{2}\right\}$ such that $\sigma(\tilde x)=\frac{1}{2}$ and $\tilde x=P(\tilde x, \sigma)(\overline \theta-\underline \theta)+\underline \theta$, and (iv) $\sigma$ is symmetric, satisfying $\sigma(\tilde x + \delta)=1-\sigma(\tilde x-\delta)$ for any $\delta \in[0, \min\{\tilde x, 1-\tilde x\}]$. 
\end{proposition}
\begin{proof}
    See \hyref{subsection:proposition:sym-qre:global}[Appendix]. 
\end{proof}

\begin{figure}[!ht]
    \centering
    \includegraphics[width=.45\linewidth]{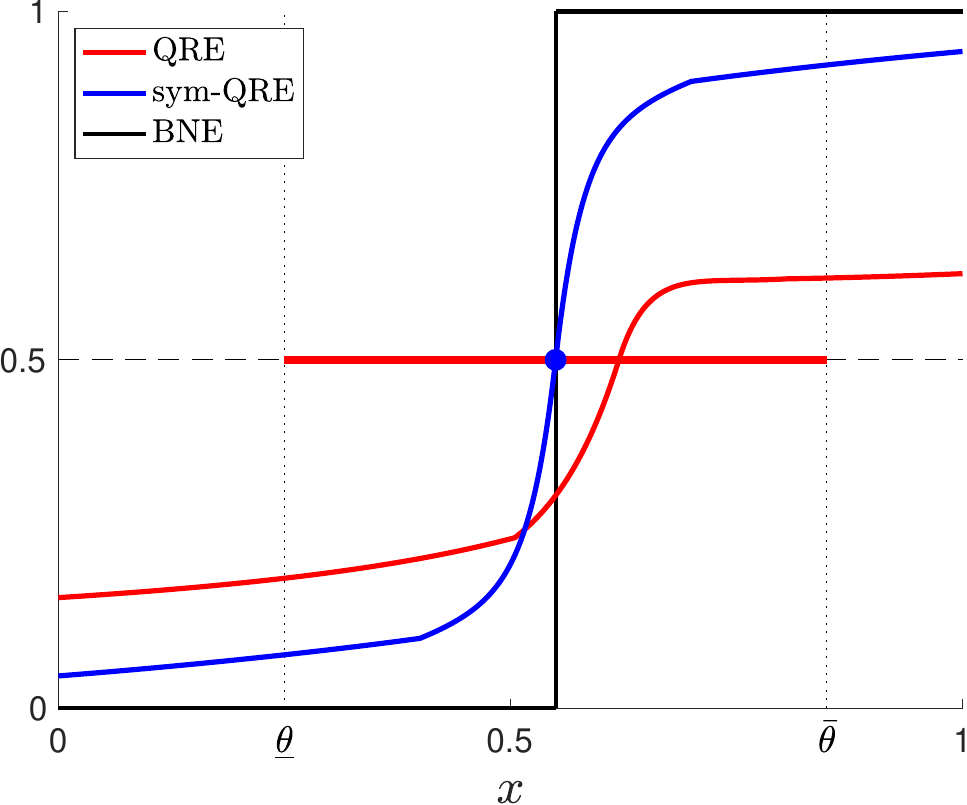}
    \begin{minipage}{1\linewidth}
        \small
        \vspace*{.5em}
        \caption{QRE and sym-QRE in the Global game}
        \label{figure:global}
        \vspace*{-1.0em}
        \singlespacing \emph{Notes}: 
        The figure depicts the NE (black) as well as illustrative QRE (red) and sym-QRE (blue), for $(k,c,\epsilon)=(0.25,0.60,0.15)$. 
        The thick horizontal line at one-half indicates the set of indifferent types under QRE; the blue circle gives the set of indifference types under sym-QRE---which is a singleton in this application.
    \end{minipage}
\end{figure}

\hyref{figure:global}[Figure] shows representative QRE (in red) and sym-QRE (in blue) for parameters $(k,c,\epsilon)=(0.25,0.60,0.15)$.  
QRE is consistent with a range of possible indifferent types, drawn as a horizontal red line. 
This particular QRE features a bias in favor abstaining:
there exists $x<\tilde x<x'$ such that $\sigma(x)<1-\sigma(x')$ and $\vert\Delta \bar u_{x}(\sigma)\vert\geq\vert\Delta \bar u_{x'}(\sigma)\vert$.
By contrast, the sym-QRE is consistent with only one indifferent type, $\tilde x=\frac{2k+c}{2}=0.55$, which coincides with that under Bayesian Nash equilibrium. 
We also see that, like the Bayesian Nash equilibrium, the sym-QRE is symmetric about this indifferent type.

\subsection{Compromise game}
\label{subsection:compromise}

The compromise game, introduced by \citet{Carrillo2009}, is a game in which two parties, privately informed of their strength, simultaneously choose whether to seek compromise or unilaterally trigger a conflict in which case the stronger player wins. 
It is a game of `two-sided adverse selection' in which the prediction of standard theory --- found via an unravelling argument --- is for all parties to choose conflict. 
The most obvious application is to war, in which case the parties are nations, but other applications include litigation, electoral debates, and firm competition.

Following \citet{Carrillo2009}, two players simultaneously decide whether to \emph{fight} (action 1) or \emph{flee} (action 0). 
Each player has a private type representing their strength $x\sim U[0,1]$, i.i.d. across players. 
If at least one player chooses to fight, the stronger player receives the high payoff of $1$ and the weaker player receives the low payoff of $0$.\footnote{
    In the measure zero event both players fight and have the same type, suppose they each receive $1$. 
} If both players flee, each receives the compromise payoff $B\in(0,1/2]$.
Let $\sigma(x)$ denote the probability that a player with strength $x$ chooses to fight.
We then have that $\bar u_x^1(\sigma)=\mathbb E[1\{x'\leq x\}]=x$, $\bar u_x^0(\sigma)=\mathbb E[\sigma(x')1\{x'\leq x\}]+B\mathbb E[1-\sigma(x')]$, and $\Delta \bar u_x(\sigma)=x-\mathbb E[\sigma(x')1\{x'\leq x\}]-B\mathbb E[1-\sigma(x')]$.

Let $G_\sigma(x):=\mathbb E[(1-\sigma(x'))1\{x'\leq x\}]/\mathbb E[1-\sigma(x')]$ denote the cumulative distribution of types, conditional on fleeing as induced by $\sigma$.
We can then write $\Delta \bar u_x(\sigma)=
(G_\sigma(x)-B)\mathbb E[1-\sigma(x')]$.
Hence, the payoff difference for type $x$ depends on $\sigma$ only through its expectation $\mathbb E[\sigma(x')]$ and the induced distribution of types $G_\sigma(x)$.

In this game, the (essentially) unique (symmetric) Bayesian Nash equilibrium is for both players to fight no matter their type: $\sigma^{BNE}=1$.\footnote{
    The type $x=0$ is indifferent and may arbitrarily mix.
} 
Clearly, there can be no Bayesian Nash equilibrium in which higher types flee and lower types fight. 
But we also cannot have a Bayesian Nash equilibrium in which higher types fight when lower types flee. 
This is because a player is only pivotal when the other player flees; and so, conditioning on the pivotal event, the flee-ers with the highest types would deviate. 
In QRE, all types will flee with some probability and so, some low types will actually prefer to flee, as fleeing is a best response to a higher type that also flees. 
In other words, noise in choice breaks the unravelling logic that underlies the Bayesian Nash equilibrium.

We now characterize QRE of the compromise game. In this game, only a weaker form of \ref{assumption:a2} and \ref{assumption:a3} hold (see \ref{section:conditions}), so we augment QRE with translation invariance \ref{assumption:s1}, which leads to the same characterization in \hyref{theorem:rqre:characterization}[Theorem].\footnote{
    The construction used in the proof of the "if" direction of \hyref{theorem:rqre:characterization}[Theorem] satisfies \ref{assumption:s1}. 
} As before, let $\tilde x(\sigma)$ denote the indifferent type and $\tilde X$ denote the set of indifferent types under QRE.
We obtain the following characterization:
\begin{proposition}
    \label{proposition:rqre:compromise}
    A strategy $\sigma \in \Sigma$ is a QRE satisfying translation invariance (\ref{assumption:s1})
    if and only if 
    (i) $\sigma$ is continuous and strictly increasing, (ii) $\sigma \in (0,1)$, and (iii) there exists a unique indifferent type $\tilde x \in \tilde X=(0,B)$ such that $\sigma(\tilde x)=\frac{1}{2}$ and $\tilde x=G_{\sigma}^{-1}(B)$. 
\end{proposition}
\begin{proof}
    See \hyref{subsection:proposition:rqre:compromise}[Appendix]. 
\end{proof}

\begin{figure}[!ht]
    \centering
    \includegraphics[width=.45\linewidth]{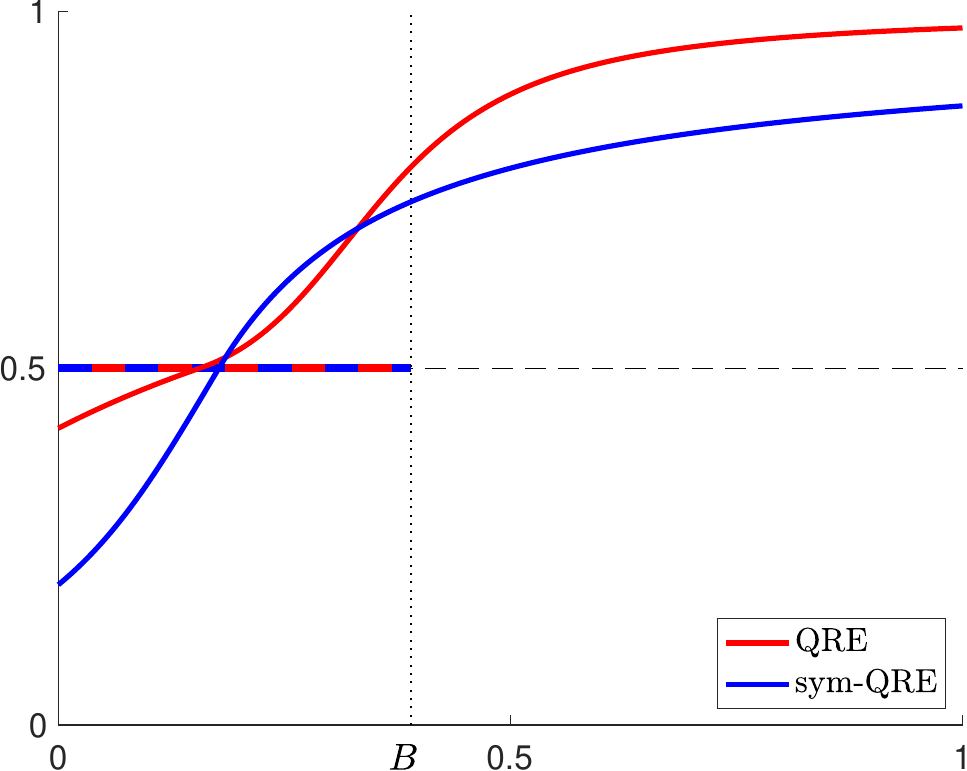}
    \begin{minipage}{1\linewidth}
        \small
        \vspace*{.5em}
        \caption{QRE and sym-QRE in the Compromise game}
        \label{figure:compromise}
        \vspace*{-1.0em}
        \singlespacing \emph{Notes}: 
        The figure depicts illustrative QRE (red) and sym-QRE (blue) for $B=0.39$. The QRE (red) is biased in favor of fighting. 
        The thick horizontal line at one-half indicates the set of indifferent types under QRE and sym-QRE, which coincide. 
    \end{minipage}
\end{figure}

The top left panel of \hyref{figure:compromise}[Figure] shows a QRE (red). 
Though it requires an argument that we develop later on, we claim that this is an example in which types are biased in favor of fighting in the sense that $\sigma(x)>1-\sigma(x')$  whenever $x<\tilde x <x'$ and $\vert\Delta\bar u_{x}(\sigma)\vert=\vert\Delta\bar u_{x'}(\sigma)\vert$.

Next, we show that imposing symmetry does not reduce the set of possible indifferent types.
\begin{proposition}
    \label{proposition:sym-qre:compromise}
    A strategy $\sigma \in \Sigma$ is a sym-QRE    
    if and only if 
    (i) $\sigma$ is continuous and strictly increasing, (ii) $\sigma \in (0,1)$, (iii) there exists a unique indifferent type $\tilde x \in \tilde S = \left(0, B\right)$ such that $\sigma(\tilde x)=\frac{1}{2}$ and $\tilde x=G_{\sigma}^{-1}(B)$, and (iv) $\sigma$ is symmetric. 
\end{proposition}
\begin{proof}
    See \hyref{subsection:proposition:sym-qre:compromise}[Appendix]. 
\end{proof}

The top right panel of \hyref{figure:compromise}[Figure] shows a sym-QRE (blue), which unlike the QRE in the left panel, satisfies symmetry.

While we have seen that symmetry does not affect the set of indifferent types, it does impose significant structure on $\sigma$.
The next result gives two necessary conditions for symmetry that may be useful in applications for identifying symmetry violations. 
\begin{corollary}
    \label{corollary:compromise}
    In any sym-QRE, 
    (1) $\sigma(\tilde x+k)<1-\sigma(\tilde x-k)$
    for any $k \in[0,\tilde x]$, 
    and $(2)$ if $B<\frac{1}{2}$,
    there exists $\overline x\in (\tilde x, 1)$ such that $\sigma(x)>1-\sigma(0)$ for all $x>\overline x$. 
\end{corollary}
Violations of these two conditions point to qualitatively different biases that may be part of QRE, but not sym-QRE. 
For instance, if \emph{(1)} is violated, some types are biased in favor of fighting. 
If (2) is violated, some types are biased in favor of fleeing. 
An example of the former is given in the top left panel of \hyref{figure:compromise}[Figure].

\begin{proof}
    See \hyref{subsection:corollary:compromise}[Appendix]. 
\end{proof}

\section{Nonparametric estimation and inference}
\label{section:empirics}

\subsection{Generic methodological guidelines}

We now show how one can leverage our results to estimate the QRE primitives and as well as test if the data is consistent with some QRE.

We will write $X\sim f$ to denote a player's type and $Y\sim\sigma(X)$ to denote their choice, where $Y=1$ with probability $\sigma(X)$ and $Y=0$ with complementary probability; we write $D:=(X,Y)$ to denote the pair.
We consider data given by the choices of $n$ individuals, where $D^n:=\{(X_i,Y_i)\}_{i=1,...,n}$ corresponds to the empirical version of $D$.

Estimation of the QRE primitive is straightforward as \hyref{theorem:recover}[Theorem] indicates how to use $\sigma$ to recover the quantal response function $Q$.\footnote{
    To be precise, after estimating $\sigma$, one associates (estimates of) payoffs $\bar u_x(\sigma)$ with (estimated) choice probabilities to obtain a point estimate of $Q$ for the restricted payoff domain $\{\bar u_x\sigma, x \in \mathcal X\}$.
}
To estimate $Q$, one simply needs any consistent nonparametric estimator $\hat \sigma_n$ of $\sigma$, such as kernel regression.

Testing if the data can be rationalized by some QRE is also simple.
Recall from \hyref{theorem:rqre:characterization}[Theorem] that, in order for $D$ to be rationalized by some QRE, it must be that (i) $\sigma(x)=\mathbb E[Y|X=x]$ is strictly increasing and continuous in $x$, (ii) $\sigma \in (0,1)$, and (iii) there is a unique $\tilde x \in \mathcal X$ such that $\Delta \bar u_{\tilde x}(\sigma)=0$ and $\sigma(\tilde x)=1/2$.
Without further restrictions, continuity and interiority of $\sigma$ are non-falsifiable properties.
However, there are valid tests for $\sigma$ being increasing --- but not strictly increasing --- and for the existence of a unique indifferent type who uniformly randomizes.

For monotonicity (i), we note that many estimators $\hat \sigma_n$ of $\sigma$ are not only consistent, but also asymptotically normal (satisfying a functional central limit theorem), e.g. nonparametric kernel regression with undersmoothed bandwidths --- see \citet{Gyorfietal2002} for a textbook reference.
One can then estimate $\sigma$ and test the null hypothesis of monotonicity of $\sigma$ by relying on standard tests, such as \citet{HallHeckman2000}, \citet{Ghosal2000}, or \citet{DelgadoEscanciano2012}.

For (iii), for simplicity, we assume that indifferent types $\tilde x$ are fully characterized by a statistic of $D$, i.e. that there is a function $T:D\mapsto T(D)\in \mathcal X$ such that $T(D)=\tilde x$ if and only $\Delta \bar u_{\tilde x}(\sigma)=0$.
In many applications of interest, such as the ones pursued in this paper, the function $T$ is easy to characterize and corresponds to natural moment conditions, which further simplifies the analysis.
For instance, in the volunteer's dilemma, $T(D)=B \mathbb E[Y_i]$ for some known parameter $B\in (1,2)$, whereas in the compromise game $T(D)=F_{X|Y=1}^{-1}(B)$, for a parameter $B\in (0,1/2]$, where $F_{X|Y=1}$ denotes the conditional distribution of $X|Y=1$ and $F_{X|Y=1}^{-1}(c):=\inf\{x\in \mathcal X\mid F_{X|Y=1}(x)>c\}$ is its quantile function.

Then, insofar as a consistent estimator $\hat T(D_n)$ of $T(D)$ is available, one can use a plug-in estimator to test if $\sigma(T(D))=1/2$ by relying on $\hat \sigma_n(\hat T(D_n))$.
Valid confidence intervals for $\sigma(T(D))$ can be obtained using bootstrap procedures \citep[see e.g.][]{HallHorowitz2013}.
Testing (iii) is then equivalent to testing if 1/2 belongs to such confidence intervals.

Naturally, this approach has the limitation that it requires knowing the payoff function $u$, since our identification results require the applicability of \hyref{theorem:rqre:characterization}[Theorems]-\ref{theorem:recover}, and the statistic $T$ depends on the payoff function $u$, since it captures the indifference condition $\Delta \bar u_{\tilde x}(\sigma)=0$.
Under expected utility, it may be possible to 
(a) normalize payoffs to be 0 and 1 when the outcomes are binary; 
(b) argue that stakes are small enough so that $u$ is close to linear;
(c) pay in probability points, a practice that is conventional in experiments (even if it is unclear whether this indeed linearizes payoffs); 
or
(d) estimate payoffs independently.
For instance, regarding (a), we have that in the compromise game players either receive 0 or $B$, and so one can without loss normalize $u(0)$ to zero and $u(B)$ to $B$, since affine transformations do not change preferences over lotteries. 
The normalization in (a) would also be without loss when allowing for individual idiosyncratic regular quantal response functions $Q_i$: we would still have a `representative' regular QRE as in Theorem 4 of \citet{Golman2011}.

\subsection{Empirical analysis of the compromise game}
We now leverage our analysis of the compromise game to nonparametrically test whether QRE is able rationalize the experimental data in \citet{Carrillo2009}.

The experiment has two variants corresponding to two values for the compromise payoff, $B\in \{.39,.50\}$, with types drawn uniformly at random from $[0,1]$, in increments of .01.
The 56 recruited subjects were students at Princeton University and played for 20 incentivized rounds with random rematching and randomly redrawn types.\footnote{
    Their experiment also includes two variants where choices are sequential; these were explained and played only after the simultaneous choice rounds were.
    We focus on the data with simultaneous choices as it matches our theoretical application from \hyref{subsection:compromise}[Section].
}

\citet{Carrillo2009} discuss the support for quantal response equilibrium based on comparing goodness-of-fit to other models.
The authors first observe that, contra Nash equilibrium, fighting rates are strictly positive, increasing in strength $x$, and decreasing in the compromise payoff $B$ --- features consistent with QRE. 
They then fit different parametric models to the data --- variations of logit QRE, Poisson-based cognitive hierarchy \citep{Camerer2004}, and cursed equilibrium \citep{Eyster2005}, among others. 
They find that, while the QRE models provide a fairly good fit, they fail to capture the tendency of subjects to ``fight with probability close to one when their strength is sufficiently high and with probability close to zero when their strength is sufficiently low.'' 
In order to capture this feature, they augment QRE with a cursedness parameter ($\alpha$-QRE) and find a statistically better fit (rejecting $\alpha$ = 0).

Importantly, the conclusions drawn in \citet{Carrillo2009} regarding QRE are entirely based on the logit functional form. 
To the extent these QRE models do not fully explain the data, a natural question then is whether QRE with a more general error structure can, in which case one need not posit additional behavioral parameters.

To answer this question, we will test whether \emph{any} QRE is able to rationalize the data.
We follow the general methodology delineated above to nonparametrically test for the adequacy of QRE to rationalize data. 

\begin{figure}[!ht]
    \centering
    \makebox[1.2\linewidth][c]{
    \begin{minipage}{1.1\linewidth}\centering
        \hspace*{-7em}\begin{subfigure}{.49\linewidth}
            \includegraphics[width=1\linewidth]{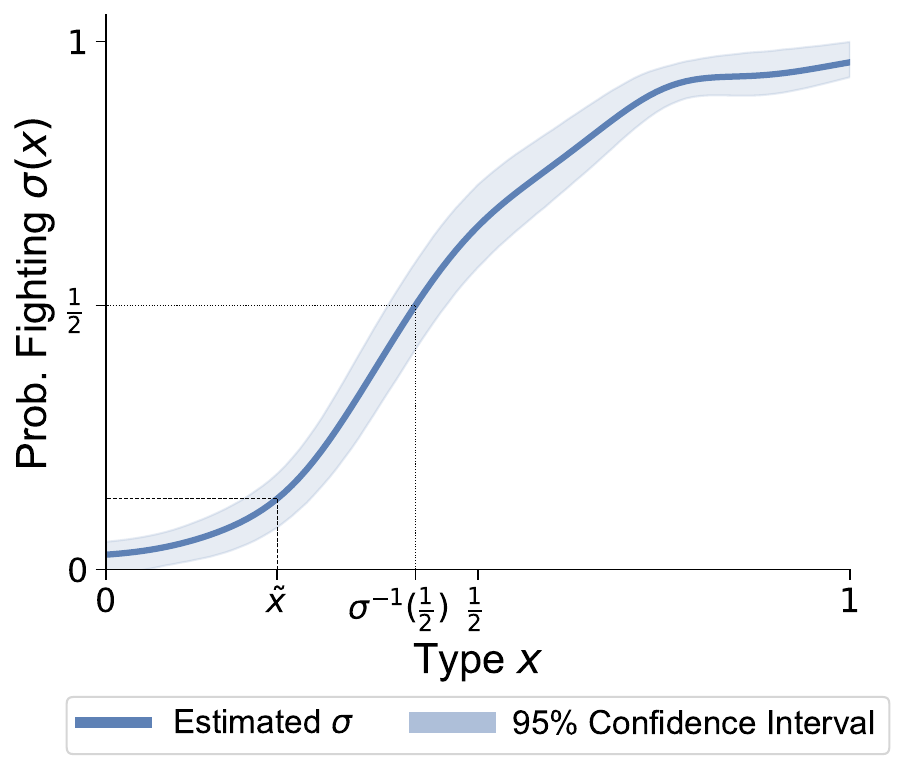}
            \caption{Compromise Payoff $B$ = .50}
            \label{figure:kernel-50}
        \end{subfigure}
        \begin{subfigure}{.49\linewidth}
            \includegraphics[width=1\linewidth]{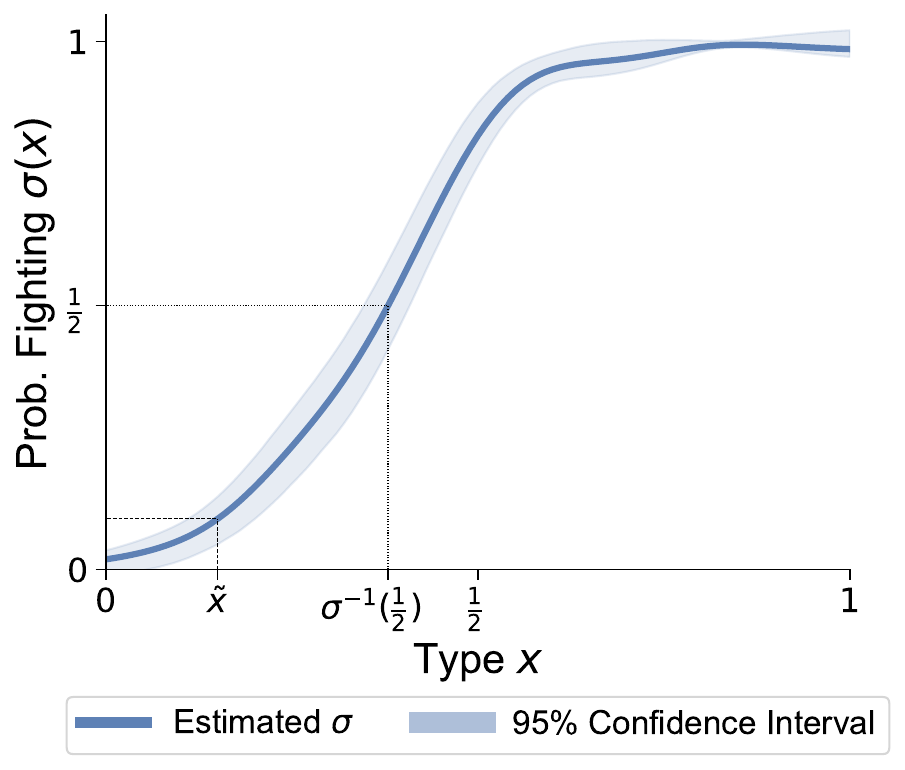}
            \caption{Compromise Payoff $B$ = .39}
            \label{figure:kernel-40}
        \end{subfigure}
    \end{minipage}
    }
    \begin{minipage}{1\linewidth}
        \small
        \vspace*{.5em}
        \caption{Estimated Choice Probability}
        \label{figure:kernel}
        \vspace*{-1.0em}
        \singlespacing \emph{Notes}: 
        The figure displays the estimated choice probability per type (line) and its 95\% bootstrap confidence interval (shaded region) for the compromise game (\hyref{subsection:compromise}[Section]).
        $\tilde x$ denotes the estimated indifferent type, $\tilde x = \hat{F}_{X|Y=1}^{-1}(B)$, whereas $\sigma^{-1}(1/2)$ is the type estimated to be uniformly randomizing.
        Point estimates were based on nonparametric kernel regression using a Gaussian kernel with a bandwidth $h=n^{-3/10} \hat s(Y)$, where $n$ refers to the number of observations and $\hat s(Y)$ the estimated standard deviation in choices.
        The data is from \citet{Carrillo2009}; we focus on simultaneous choice treatments.
    \end{minipage}
\end{figure}

Specifically, by observing types and actions, we nonparametrically estimate $\sigma$ via kernel regression, using a Gaussian kernel with a bandwidth $h_n=n^{-3/10}\hat s(Y)$, where $\hat s(Y)$ is an unbiased estimate of the standard deviation of $Y$.\footnote{
    The chosen bandwidth is based on Silverman's rule-of-thumb, but undersmoothed so as to guarantee consistency.
} 
\hyref{figure:kernel}[Figure] shows the estimated function for each of the two values of $B$, a clearly monotone function, where the type uniformly randomizing is in the admissible set for QRE, $\sigma^{-1}(1/2)\in (0,B)$.
However, as shown, the empirically indifferent type $\tilde x$ does not coincide with the the type estimated to be uniformly randomizing, $\sigma^{-1}(1/2)$. 
In particular, we find that $\tilde x$ is significantly lower than $\sigma^{-1}(1/2)$, which already suggests that the data cannot be rationalized by any QRE, in line with our characterization results.
In other words, there seems to be a bias toward fleeing, which suggests subjects may be underestimating the frequencies with which lower types flee and/or higher types fight.

Then, we test whether $\sigma$ is increasing; as mentioned above, continuity of $\sigma$, \emph{strict} monotonicity, and interiority are not falsifiable.
We test monotonicity using the procedure developed by \citet[Section 3]{DelgadoEscanciano2012}; we do not find any significant violations of monotonicity, even for high significance levels (e.g. $\alpha =0.80$).

Finally, we test if the empirically indifferent type $\tilde x$ uniformly randomizes, $\sigma(\tilde x)=1/2$.
As discussed above, $\Delta \bar u_{\tilde x}(\sigma)=0\Longleftrightarrow \tilde x = F_{X|Y=1}^{-1}(B)$, where $F_{X|Y=1}^{-1}$ is the quantile function associated with the distribution of $X|Y=1$.
Thus, we consistently estimate $\tilde x$ via standard quantile estimation, and obtain a confidence interval for $\sigma(\tilde x)$ based on bootstrapping.
The resulting confidence intervals are shown in \hyref{table:table-compromise-stat-ci}[Table], which decidedly reject QRE --- and therefore sym-QRE as well --- since we can reject the hypothesis that the indifferent type uniformly randomizes.

\citet{Carrillo2009} find that logit QRE does not fully explain their data, which leads them to consider other behavioral models. 
We view our result as strengthening this conclusion: since a general form of QRE is rejected, one must step outside of the QRE family in order to fully explain the data.

\begin{table}[!ht]
    \setstretch{1.1}
	\centering
	\small
	
\begin{tabular}{@{\extracolsep{4pt}}ccc@{}}
\hline\hline
& \multicolumn{2}{c}{Compromise Payoff}  \\
& \multicolumn{1}{c}{.50}  &  \multicolumn{1}{c}{.39} \\
\cline{2-2} \cline{3-3}
& (1) & (2) \\
\hline
$\tilde x$ & .230 & .150 \\
$\sigma^{-1}(1/2)$ & .416 & .379 \\
$\sigma(\tilde x)$ & .134 & .096 \\
\,\,95\% CI & (.080, .182) & (.048 , .147) \\
\,\,99\% CI & (.069 , .189) & (.038 , .144) \\
\hline\hline
\end{tabular}
    \begin{minipage}{1\linewidth}
        \small
        \vspace*{.5em}
        \caption{Testing QRE}
        \label{table:table-compromise-stat-ci}
        \vspace*{-1.5em}
        \singlespacing \emph{Notes}: 
        This table tests the adequacy of QRE in rationalizing the data for the compromise game (\hyref{subsection:compromise}[Section]) from \citet{Carrillo2009}; we focus on simultaneous choice treatments.
        $\tilde x$ denotes the estimated indifference type, $\sigma^{-1}(1/2)$ the type estimated to be uniformly randomizing, $\sigma(\tilde x)$ the probability with which the indifferent type randomizes, and it provides bootstrapped confidence intervals for $\sigma(\tilde x)$ based on 10,000 replications.
        By \hyref{proposition:rqre:compromise}[Proposition], $\sigma(\tilde x)=1/2$, that is, $\tilde x$ should equal $\sigma^{-1}(1/2)$.
    \end{minipage}
\end{table}

\section{Conclusion}
\label{section:conclusion}

Quantal response equilibrium (QRE) explains many of the well-known deviations from Nash equilibrium (NE) observed in the lab. 
It should also be regarded as an important theoretical benchmark in that it deviates in a minimal way from NE. 
Nevertheless, its influence in theoretical applications is limited, perhaps due to concerns over its tractability. 
Recent work, focusing on finite games, has made great strides by analyzing more general non-parametric forms. 
This has opened up the potential for richer applications and new ways of organizing experimental data. 

In this paper, we provide analogous results for a common class of infinite games, those with binary actions and a continuum of types. 
Specifically, under a weak monotonicity condition on payoffs, we show that the full set of QRE is characterized by three simple conditions on choice probabilities: continuity, monotonicity, and uniform mixing of indifferent types.
Further, we show how to recover the quantal response function from observable choices and types.
We then apply our results to characterize QRE in a number of classic games and obtain sharp predictions.
We conclude by illustrating the usefulness of our characterization in developing nonparametric tests of QRE.
We believe that these results will inform both theoretical and empirical research, reducing reliance on parametric assumptions.

\section{References}
\setlength\bibhang{0pt}
\setlength{\bibsep}{0em plus 0ex}
\bibliographystyle{econ-aea}
\bibliography{continuous-qre.bib}

\newpage

\setcounter{section}{0}
\renewcommand{\thesection}{Appendix \Alph{section}}
\renewcommand{\thesubsection}{\Alph{section}.\arabic{subsection}}
\section{Omitted Proofs}
\label{section:appendix}

\subsection{Proof of \hyref{proposition:BNE}[Proposition]}
\label{subsection:proposition:BNE}
\begin{proof}
    Fix any Bayesian Nash equilibrium $\sigma$.
    First, we show that any Bayesian Nash equilibrium $\sigma$ is increasing.
    Suppose for the purpose of contradiction that $\sigma$ is not increasing. 
    Then there are $x<x'$ such that $\sigma(x)>\sigma(x')$.
    By \ref{assumption:a2}, there are $\hat x\neq \hat x'$ such that (i) $\sigma(\hat x)> \sigma(\hat x')$, and (ii) $\Delta \bar u_{\hat x}(\sigma)<\Delta \bar u_{\hat x'}(\sigma)$.
    The latter condition implies that action 1 is the unique best response for type $\hat x'$, and $\sigma(\hat x')=1$, or action 0 is the unique best response for type $\hat x$, and $\sigma(\hat x) =0$.
    Then, $\sigma(\hat x)> \sigma(\hat x')=1$ or $0=\sigma(\hat x)> \sigma(\hat x')$, a contradiction.

    Now we show that any Bayesian Nash equilibrium must be in threshold strategies.
    For the purpose of contradiction, suppose that $\sigma$ is an equilibrium not in threshold strategies. 
    As it must be increasing, then we must then have $x,x'$ such that $x'>x$ and $0<\sigma(x)\leq \sigma(x')<1$.
    Then, both types $x$ and $x'$ must be indifferent between actions 0 and 1, i.e. $\Delta \bar{u}_x(\sigma)=\Delta \bar u_{x'}(\sigma)=0$, which contradicts \ref{assumption:a3}.

    We now show existence of such equilibria.
    Note that, for any $x\in \mathcal X$, $1\{\bullet\geq x\}\in \Sigma$ is a threshold strategy.
    By \ref{assumption:a1}, $\Delta \bar u_x(1\{\bullet\geq x\})$ is continuous in $x$. 
    By \ref{assumption:a3}, $\Delta \bar u_x(1\{\bullet\geq x\})\leq 0$ implies $\Delta \bar u_{x'}(1\{\bullet\geq x\})<0$ for all $0\leq x'<x$, and 
    $\Delta \bar u_x(1\{\bullet\geq x\})\geq 0$ implies $\Delta \bar u_{x'}(1\{\bullet\geq x\})>0$ for all $1\geq x'>x$.

    Therefore, if $\Delta \bar u_{0}(1\{\bullet\geq 0\})\geq 0$ or 
    $\Delta \bar u_{1}(1\{\bullet\geq 1\})\leq 0$, 
    then $\sigma =0$ or $\sigma =1$, respectively, are Bayesian Nash equilibria.
    Otherwise, by continuity, there is a $\tilde x \in \mathcal X$ such that $\Delta \bar{u}_{\tilde x}(1\{\bullet\geq \tilde x\})=0$, and $1\{\bullet\geq \tilde x\}$ is a Bayesian Nash equilibrium.
\end{proof}

\subsection{Proof of \hyref{lemma:existence}[Lemma]}
\label{subsection:lemma:existence}
\begin{proof}
    We first derive properties that must be satisfied by any QRE.

    Take any $\sigma\in \Sigma$ satisfying $\sigma=q(\sigma)$.
    From \ref{assumption:r1}, $\sigma(x)\in (0,1)$ for any $x\in \mathcal X$.
    From \ref{assumption:r4}, we have that $\sigma(x)=1/2\Longleftrightarrow \Delta \bar u_x(\sigma)=0$. 
    We first show that any such $\sigma$ must be continuous.
    Fix any $x\in \mathcal X$ and take an arbitrary sequence $\{x_n\}_{n \in \mathbb N}\subseteq \mathcal X$ converging to $x$.
    Then, by \ref{assumption:a1}, $\bar u_{x_n}(\sigma)\to \bar u_x(\sigma)$ which, by \ref{assumption:r2} implies $Q(\bar u_{x_n}(\sigma))\to Q(\bar u_x(\sigma))$ and thus $\sigma(x_n)=q(\sigma)(x_n)=Q(\bar u_{x_n}(\sigma))\to Q(\bar u_{x}(\sigma))=\sigma(x)$.

    We now prove that any such $\sigma$ must be increasing.
    Suppose for the purpose of contradiction that there are $x<x'$ such that $\sigma(x)> \sigma(x')$.
    Then, by \ref{assumption:a2}, there are $\hat x, \hat x'$ such that $\sigma(\hat x)> \sigma(\hat x')$ and $\bar u_{\hat x}^1(\sigma)\leq \bar u_{\hat x'}^1(\sigma)$ and $\bar u_{\hat x}^0(\sigma)\geq \bar u_{\hat x'}^0(\sigma)$ with at least one of the payoff inequalities being strict.
    From \ref{assumption:r3}, $\sigma(\hat x) = q(\sigma)(\hat x) = Q(\bar u_{\hat x}(\sigma))<Q(\bar u_{\hat x'}(\sigma)) = q(\sigma)(\hat x') = \sigma(\hat x')$, a contradiction.

    Finally, we show that any QRE $\sigma$ must be strictly increasing.
    Suppose $\sigma=q(\sigma)$ is increasing but not strictly increasing.
    Then, there are $x,x'$ such that $x<x'$ and $\sigma(x)=\sigma(x')$. 
    By \ref{assumption:a3}, $\bar u_{x}^1(\sigma)-\bar u_{x'}^1(\sigma)\leq 0\leq \bar u_{x}^0(\sigma)-\bar u_{x'}^0(\sigma)$, with at least one of the inequalities strict.
    Again from \ref{assumption:r3}, $\sigma(x) = q(\sigma)(x) = Q(\bar u_{x}(\sigma))<Q(\bar u_{x'}(\sigma)) = q(\sigma)(x') = \sigma(x')$, a contradiction.

    We then note that the image of $q$, $q(\Sigma)$, is a subset of the space of continuous increasing functions mapping from compact set $\mathcal X$ to $[0,1]$, denoted $\mathcal S$, which is compact with respect to the bounded variation norm, $\|\cdot\|_{BV}$.\footnote{
        That is, $\|\sigma\|_{BV} := \|\sigma\|_{L^1}+V(\sigma)$, where $V(\sigma)$ denotes the total variation of $\sigma \in \mathcal S$.
    }
    We now want to show that $q$ (restricted to $\mathcal S$, the relevant domain) is continuous with respect to $\|\cdot \|_{BV}$.
    Let $\|\bar u\|_\infty:=\max_{x \in \mathcal X, \sigma \in \mathcal S} \|\bar u_x(\sigma)\|_\infty$, which is well-defined by Weierstrass extremum theorem and \ref{assumption:a1}.
    Let $\mathcal U:=[-\|\bar u\|_\infty,\|\bar u\|_\infty]^2\subset \mathbb R^2$, which is a compact superset of the domain of expected payoffs when restricting opponents' (symmetric) strategies to $\mathcal S$.
    Restricting $Q$ to $\mathcal U$ renders it uniformly continuous by Heine-Cantor's theorem.
    Take any sequence $\{\sigma_n\}_n \subseteq \mathcal S$ such that $\|\sigma_n-\sigma\|_{BV}\to 0$.
    Since $\|\sigma_n-\sigma\|_{BV}\to 0 \Longrightarrow \|\sigma_n-\sigma\|_{L^1}\to 0$, then, by Berge's theorem of the maximum, $\max_{x \in \mathcal X}\|\bar u_x(\sigma_n)-\bar u_x(\sigma)\|_\infty$ is continuous in $\sigma_n$ and so converges to zero.
    This implies that for every $\epsilon>0$ there exists $N<\infty$ such that for all $n>N$, and all $x\in \mathcal X$, $\|\bar u_x(\sigma_n)-\bar u_x(\sigma)\|_\infty<\epsilon$.
    Combining this with uniform continuity of $Q$ when restricted to the relevant domain, we obtain that for every $\epsilon>0$ there exists $N<\infty$ such that for all $n>N$, and all $x\in \mathcal X$, $|q(\sigma_n)(x)-q(\sigma)(x)|=|Q(\bar u_x(\sigma_n))-Q(\bar u_x(\sigma))|<\epsilon$, and therefore $q(\sigma_n)$ converges to $q(\sigma)$.

    Finally, we observe that $q(\Sigma)\subseteq \mathcal S$ and thus $q(\mathcal S)\subseteq \mathcal S$, and that $\mathcal S$ is in turn a subset of the space of functions with bounded variation defined on $\mathcal X$, $BV(\mathcal X):=\{f \in [0,1]^{\mathcal X}\mid f\text{ is continuous and } V(f)<\infty\}$, itself a Banach space with respect to $\|\cdot\|_{BV}$.
    Since $\mathcal S$ is compact with respect to $\|\cdot \|_{BV}$ and convex, by Schauder's fixed-point theorem, a fixed point $\sigma = q(\sigma)$ exists.
\end{proof}

\subsection{Proof of \hyref{proposition:rqre:volunteer}[Proposition]}
\label{subsection:proposition:rqre:volunteer}

\begin{proof}
    Fix a QRE $\sigma$. 
    Note that for the indifferent type, $\tilde x\equiv\tilde x(\sigma)$, $\Delta \bar u_{\tilde x}(\sigma)=0 \Longleftrightarrow 
    \tilde x/B=\mathbb E[\sigma(x')] =  \mathbb P(x \leq \tilde x)\mathbb E[\sigma(x')\mid x \leq \tilde x] + \mathbb P(x > \tilde x)\mathbb E[\sigma(x')\mid x > \tilde x]$.
    As $\sigma(\tilde x)=1/2$ and $\sigma$ is strictly increasing, then
    $\mathbb E[\sigma(x')\mid x \leq \tilde x]< \frac{1}{2} < \mathbb E[\sigma(x')\mid x > \tilde x]$, implying
    $\mathbb P(x > \tilde x)\frac{1}{2}<\tilde x/B<\mathbb P(x \leq \tilde x)\frac{1}{2}+\mathbb P(x > \tilde x)
    \Longleftrightarrow
    (1-\tilde x)/2<\tilde x/B<(1-\tilde x/2)
    \Longleftrightarrow
    B/(2+B)<\tilde x<2B/(2+B)$.

    Now fix $\tilde x \in (B/(2+B), 2B/(2+B))$. 
    Let $\hat \sigma\in [0,1]^{\mathcal X}$ be such that $\hat \sigma(x):=\alpha$ for $x\in [0,\tilde x)$, $\hat \sigma(\tilde x)=1/2$, and $\hat \sigma(x):=\beta$ for $x\in (\tilde x,1]$, where $0<\alpha<1/2<\beta<1$.
    From $\Delta \bar u_{\tilde x}(\hat \sigma)=0$, we then have that $\tilde x = \frac {\beta B} {1 + \beta B - \alpha B}$, with the right-hand side continuous and increasing in $\alpha$ and $\beta$, attaining $B/(2+B)$ when $\alpha=0, \,\beta=1/2$, and attaining $2 B/(2+B)$ when $\alpha=1/2, \,\beta=1$.
    For any $\alpha,\beta:0<\alpha<1/2<\beta<1$, one can then get a continuous and strictly increasing $\sigma \in (0,1)$ such that $\sigma(\tilde x)=\hat \sigma(\tilde x)=1/2$, $\mathbb E[\sigma(x')|x'\leq \tilde x]=\alpha$, and $\mathbb E[\sigma(x')|x'> \tilde x]=\beta$.
    Any such $\sigma$ will by \hyref{theorem:rqre:characterization}[Theorem] constitute a QRE.
\end{proof}

\subsection{Proof of \hyref{proposition:sym-qre:volunteer}[Proposition]}
\label{subsection:proposition:sym-qre:volunteer}
\begin{proof}
    Fix a sym-QRE $\sigma$.
    Since by \hyref{theorem:sym-qre:characterization}[Theorem] $\sigma$ is symmetric, 
    and $\Delta \bar u_{\tilde x-\delta}(\sigma)=
    \tilde x-\delta-B \mathbb E[\sigma(x')]
    =
    -\delta
    =
    -
    \Delta \bar u_{\tilde x+\delta}(\sigma)
    $, then $\sigma(\tilde x + \delta)=1-\sigma(\tilde x - \delta)$ for $\delta\leq \min\{\tilde x,1-\tilde x\}$.
    
    We now show that $\tilde x\geq 1/2$. 
    Suppose not; then $\tilde x<1/2$ and 
    \begin{align*}
        \tilde x/B
        &=
        \mathbb E[\sigma(x')]
        =
        \int_0^{\tilde x}\sigma(x')\diff x' 
        + \int_{\tilde x}^{2 \tilde x}\sigma(x')\diff x' 
        + \int_{2\tilde x}^1\sigma(x')\diff x'
        \\
        &=
        \int_0^{\tilde x}\sigma(\tilde x-x')\diff x'
        +
        \int_{0}^{\tilde x}\sigma(\tilde x+x')\diff x' 
        +
        \int_{2\tilde x}^1\sigma(x')\diff x'
        \\
        &=
        \tilde x-
        \int_0^{\tilde x}\sigma(\tilde x+x')\diff x'
        +
        \int_{0}^{\tilde x}\sigma(\tilde x+x')\diff x' 
        +
        \int_{2\tilde x}^1\sigma(x')\diff x'
        =
        \tilde x
        +
        \int_{2\tilde x}^1\sigma(x')\diff x'
        \\
        &\Longleftrightarrow
        \tilde x(B-1)+B\int_{2\tilde x}^1\sigma(x')\diff x'
        =0,
    \end{align*}

    a contradiction.
    Hence, $\tilde x\geq 1/2$, from which we have
    \begin{align*}
        \tilde x/B
        &=
        \mathbb E[\sigma(x')]
        =
        \int_0^{2\tilde x-1}\sigma(x')\diff x' 
        + \int_{0}^{1-\tilde x}\sigma(\tilde x - x')\diff x' 
        + \int_{0}^{1-\tilde x}\sigma(\tilde x +x')\diff x'
        =
        \int_0^{2\tilde x-1}\sigma(x')\diff x' 
        + 1-\tilde x
        \\
        \Longleftrightarrow
        &
        B \int_0^{2\tilde x-1}\sigma(x')\diff x' 
        + B(1-\tilde x)
        -\tilde x=0.
    \end{align*}
    Since $\sigma(x')\in (0,1/2)$ for $x'\in [0,\tilde x)$, we have 
    \begin{align*}
        &0=
        B \int_0^{2\tilde x-1}\sigma(x')\diff x' 
        + B(1-\tilde x)
        -\tilde x
        <
        B \frac{1}{2}(2\tilde x-1)
        + B(1-\tilde x)
        -\tilde x
        \\
        \Longrightarrow 
        & 
        B (2\tilde x-1)
        + B(2-2\tilde x)
        -2\tilde x
        =
        B
        -2\tilde x
        >0
        \Longrightarrow
        \tilde x<\frac{B}{2},
    \end{align*}
    as well as 
    \begin{align*}
        0=
        B \int_0^{2\tilde x-1}\sigma(x')\diff x' 
        + B(1-\tilde x)
        -\tilde x
        >
        B(1-\tilde x)
        -\tilde x
        \Longrightarrow 
        \tilde x>\frac{B}{1+B}.
    \end{align*}

    Now take $\tilde x \in (B/(1+B),B/2)$.
    Let $\hat \sigma\in [0,1]^{\mathcal X}$ be such that 
    $\hat \sigma(x)=\delta$ for $x\leq 2\tilde x -1 $,
    $\hat \sigma(x)=\alpha$ for $x\in (2\tilde x -1,\tilde x) $,
    $\hat \sigma(\tilde x)=1/2$, and 
    $\hat \sigma(x)=1-\alpha$ for $x>\tilde x$, for some $0<\delta<\alpha<1/2$.
    Then $\hat \sigma$ is symmetric, increasing, takes values in $(0,1)$, and 1/2 at $\tilde x$.
    From $\Delta \bar u_{\tilde x}(\hat \sigma)=0$ we obtain that 
    $\tilde x =\frac{B(1-\delta)}{1+B-2\delta B}$, which is increasing in $\delta\in (0,1/2)$, attaining $B/(1+B)$ when $\delta = 0$ and $B/2$ when $\delta =1/2$. 
    For any $\alpha,\delta:0<\delta<\alpha<1/2$, one can then get a continuous and strictly increasing $\sigma \in (0,1)$ such that $\sigma(\tilde x)=\hat \sigma(\tilde x)=1/2$, $\mathbb E[\sigma(x')|x'\leq 2\tilde x-1]=\delta$, and $\mathbb E[\sigma(x')|x'> \tilde x]=1-\alpha$.
    Any such $\sigma$ will by \hyref{theorem:sym-qre:characterization}[Theorem] constitute a sym-QRE.
\end{proof}

\subsection{Proof of \hyref{lemma:global:failure}[Lemma]}
\label{subsection:lemma:global:failure}

\begin{proof}
    Fix any continuous and strictly increasing QRE $\sigma$.
    Note that $\Delta\bar u_x^0(\sigma)$ is then strictly increasing in $x$.
    From \ref{assumption:r4}, $\sigma(\tilde x)=1/2\Longleftrightarrow \Delta \bar u_{\tilde x}(\sigma)=0$.

    (1) For any $x<\underline \theta$, $\Delta \bar u_x(\sigma)<0$, and 
    $x>\overline \theta$, $\Delta \bar u_x(\sigma)>0$.
    This implies that $\tilde x:\Delta \bar u_{\tilde x}(\sigma)=0$ is such that $\tilde x \in [\underline \theta,\overline \theta]$.
    Since, at any QRE, $\sigma(x)\in (0,1)$, $\tilde x \in (\underline \theta,\overline \theta)$.
    (2) $\Delta \bar u_{\tilde x}(\sigma)=\tilde x - k - c P(\tilde x, \sigma) =0 \Longleftrightarrow P(\tilde x, \sigma)=\frac{\tilde x- \underline \theta}{\overline \theta-\underline \theta}$.
    (3) Since $P(\tilde x, \sigma)=\frac{\tilde x- \underline \theta}{\overline \theta-\underline \theta}\in (0,1)$, we have 
    $\frac{\tilde x- \underline \theta}{\overline \theta-\underline \theta}=P(\tilde x, \sigma)=\frac{\theta^*(\sigma)-\tilde x+\epsilon}{2\epsilon}$, from which derives (3).
    (4) is obtained directly from the definition of $\Delta \bar u_x(\sigma)$ and $P(x,\sigma)$.
\end{proof}

\subsection{Proof of \hyref{proposition:unique:global}[Proposition]}
\label{subsection:proposition:unique:global}

\begin{proof}
    Consider $\sigma\neq\sigma'$ as two candidate QRE for the same $Q$ satisfying \ref{assumption:r1}-\ref{assumption:r4}. 
    Assume that $\sigma$ and $\sigma'$ are continuous and strictly increasing. We now derive a contradiction. 

    In what follows, we let $\tilde x$ and $\tilde x'$ be the indifferent types under $\sigma$ and $\sigma'$, respectively. 
    Further, let $\theta^*=\theta^*(\sigma)$ and ${\theta^*}'=\theta^*(\sigma')$ be the corresponding thresholds.
    Without loss of generality, we suppose $\tilde x\leq \tilde x'$.
    Immediately, we have that $\tilde x<\tilde x'$, since the indifferent type pins down all payoffs, and therefore $\tilde x=\tilde x' \Longrightarrow \sigma=\sigma'$.

   From \hyref{lemma:global:failure}[Lemma], ${\theta^*}'>\theta^*$ and $P(x,\sigma')\geq P(x,\sigma)$, with a strict inequality on $({\theta^*}-\epsilon,{\theta^*}'+\epsilon)$, and with equality elsewhere. It follows that $\bar u_x^1(\sigma) \geq \bar u_x^1(\sigma')$, with a strict inequality on $({\theta^*}-\epsilon,{\theta^*}'+\epsilon)$, and with equality elsewhere. Noting that $\bar u_x^0(\sigma)=\bar u_x^0(\sigma')=k$ for all $x$, it must be that $\sigma=\sigma'$ on $[0,{\theta^*}-\epsilon]\cup[{\theta^*}'+\epsilon,1]$ and $\sigma>\sigma'$ on $({\theta^*}-\epsilon,{\theta^*}'+\epsilon)$.

    Since $\tilde x<\tilde x'$, we have that $P(\tilde x,\sigma)<P(\tilde x',\sigma')$.
    Note that, by (3) in \hyref{lemma:global:failure}[Lemma], $\theta^*-\tilde x=\tilde x\frac{2\epsilon}{\overline \theta-\underline \theta}+\epsilon\left(\frac{\overline \theta+\underline \theta}{\overline \theta-\underline \theta}\right)$, which is strictly increasing in $\tilde x$.
    But then, by (4) in \hyref{lemma:global:failure}[Lemma] (and recalling that $\Delta \bar u_{\tilde x}(\sigma)=\Delta \bar u_{\tilde x'}(\sigma')=0$), we have that 
    $\Delta\bar u_{\tilde x'+\delta}(\sigma')
    =\tilde x' +\delta + \frac{\overline \theta-\underline \theta}{2\epsilon}\delta
    $
    for $\delta:0< \delta \leq {\theta^*}'-\tilde x'+\epsilon$
    and
    $\Delta \bar u_{\tilde x+\delta}(\sigma)
    =\tilde x +\delta + \frac{\overline \theta-\underline \theta}{2\epsilon}\delta
    $
    for $\delta:0< \delta \leq \theta^*-\tilde x+\epsilon$.
    Then, $\Delta \bar u_{\tilde x'+\delta}(\sigma')-\Delta \bar u_{\tilde x+\delta}(\sigma)=\tilde x'-\tilde x>0$ for any $\delta \in (0,\theta^*-\tilde x+\epsilon)$.
    We know that $\Delta \bar u_{\tilde x'+\delta}(\sigma')-\Delta \bar u_{\tilde x+\delta}(\sigma)$ is continuous for all $\delta \in [-\tilde x,1-\tilde x']$ and that, at $\delta=0$, this difference is zero, which is a contradiction.

    Since (1) there is a unique strictly increasing and continuous QRE $\sigma$, (2) \hyref{lemma:global}[Lemma] establishes that the set of QRE is a complete lattice, and that (3) the largest and smallest QRE are strictly increasing and continuous, it follows that there is a unique QRE, which is strictly increasing and continuous.
\end{proof}

Let $\mathcal F:=[0,1]^{\mathcal X}$ and write $f\trianglerighteq g \Longleftrightarrow f(x)\geq g(x)$ $\forall x\in \mathcal X$.

\begin{lemma} 
    \label{lemma:global}
    The set of QRE is a complete lattice with respect to the partial order $\trianglerighteq$, and the largest and smallest QRE are continuous and strictly increasing.
\end{lemma}

\begin{proof}
    Note that $(\mathcal F,\trianglerighteq)$ is a complete lattice.
    Recall $q(\sigma):=({Q(\bar u_x(\sigma))}_{x\in \mathcal X})$ is a self-map on $\mathcal F$.
    Moreover, for any $\sigma\trianglerighteq \sigma'$, $\bar u_x^0(\sigma)=\bar u_x^0(\sigma')=\underline \theta$, while 
    $\bar u_x^1(\sigma) =x-\underline \theta -(\overline \theta -\underline \theta)P(x,\sigma)\geq x-\underline \theta -(\overline \theta -\underline \theta)P(x,\sigma')$, since $P(x,\sigma)=\int_{x-\epsilon}^{x+\epsilon}\frac{1}{2\epsilon}1\{\bar \sigma(\theta)\leq 1/2\}\diff \theta$ is decreasing in $\sigma$ with respect to $\trianglerighteq$.
    Under \ref{assumption:r3}, $q$ is then a monotone operator, in that $\sigma\trianglerighteq \sigma' \Longrightarrow q(\sigma)\trianglerighteq q(\sigma')$.
    Hence, by Tarski's fixed-point theorem the set of QRE is nonempty and forms a complete lattice $q(\sigma)=\sigma$.

    Now suppose that $\sigma$ is increasing in $x$.
    Then, $\bar \sigma$ is increasing, $P(x,\sigma)$ is decreasing in $x$, and so $\bar u_x^1(\sigma)$ is strictly increasing in $x$, finally implying, by \ref{assumption:r3}, that $q(\sigma)$ is strictly increasing in $x$.
    Additionally, noting that $q$ is continuous in $\sigma$, we have that $\limsup_n q^n(1)$ and $\liminf_n q^n(0)$ correspond to the largest and smallest QRE, where $q^n$ is the $n$-fold composition of $q$ with itself.
    Since $\sigma =1$ and $\sigma=0$ are increasing strategies, so are $\limsup_n q^n(1)$ and $\liminf_n q^n(0)$.

    Now, we note that 
    \begin{align*}
        \left|
            P(x,\sigma) - P(x',\sigma)
        \right|
        =
        \left|
            \int_{x-\epsilon}^{x-\epsilon}\frac{1}{2\epsilon}1\{\bar \sigma(\theta)\leq 1/2\}\diff \theta
            -
            \int_{x'+\epsilon}^{x'+\epsilon}\frac{1}{2\epsilon}1\{\bar \sigma(\theta)\leq 1/2\}\diff \theta
        \right|
        \leq \frac{1}{\epsilon}|x-x'|,
    \end{align*}
    and so $P(x,\sigma)$ is Lipschitz-continuous in $x$.
    Therefore, $\bar u_x (\sigma)$ is continuous in $x$ and so, for any $Q$ satisfying \ref{assumption:r2}, $Q(\bar u_x (\sigma))$ is continuous in $x$.
    Hence, any QRE $\sigma$ is continuous.
\end{proof}

\subsection{Proof of \hyref{proposition:rqre:global}[Proposition]}
\label{subsection:proposition:rqre:global}
\begin{proof}
    Fix a QRE $\sigma$. As from {(1)} in \hyref{lemma:global:failure}[Lemma], $\tilde x \in (\underline \theta,\overline \theta)$.

    Now fix $\tilde x\in(\underline \theta,\overline \theta)$. 
    From (3) in \hyref{lemma:global:failure}[Lemma], the corresponding threshold
    $\theta^*(\sigma)\in(\tilde x-\epsilon,\tilde x+\epsilon)$
    required to support it as the indifferent type is uniquely defined. 
    Conversely, any $\sigma$ that is strictly increasing with $\sigma(\tilde x)=\frac{1}{2}$ such that 
    $\frac{1}{2\epsilon}\int_{\theta^*(\sigma,\epsilon)-\epsilon}^{\theta^*(\sigma,\epsilon)+\epsilon}\sigma(x')\diff x'=\frac{1}{2}$
    will deliver $\tilde x$ as the indifferent type. 
    
    Take $\hat{\sigma}(x)=\alpha$ for $x\in [0,\tilde x)$ and $\hat \sigma(x)=\beta$ for $x\in [\tilde x,1]$, where $0<\alpha<1/2<\beta<1$ are such that $\frac{1}{2\epsilon}\int_{\tilde \theta-\epsilon}^{\tilde \theta+\epsilon}\hat{\sigma}(x')\diff x'=\frac{1}{2}$, for some arbitrary $\tilde \theta \in (\tilde x-\epsilon,\tilde x+\epsilon)$.
    
    Consider the case that $\tilde \theta<\tilde x$ (the case that $\tilde \theta\geq\tilde x$ is similar). 
    In this case, $\frac{1}{2\epsilon}\int_{\tilde \theta-\epsilon}^{\tilde \theta+\epsilon}\hat{\sigma}(x')\diff x'=
    \lambda \alpha+(1-\lambda)\beta
    $, 
    where $\lambda = \frac{\epsilon+\tilde x-\tilde \theta}{2\epsilon}\in (1/2,1)$.
    Constants $\alpha,\beta$ can always be chosen so that this expression equals $1/2$, since it is a linear combination of
    $\alpha$ and $\beta$, and $0<\alpha<1/2<\beta<1$.
    As $\tilde \theta$ was chosen arbitrarily in $(\tilde x-\epsilon,\tilde x+\epsilon)$, we can pick $\alpha,\beta$ so that the above holds for $\tilde \theta = \theta^*(\sigma)$. 
    Furthermore, $\hat \sigma$ can be approximated arbitrarily well
    by a continuous, strictly increasing $\sigma$ satisfying $\sigma(\tilde x)=1/2$. 
    Any such $\sigma$ will, by \hyref{theorem:rqre:characterization}[Theorem], constitute a QRE. 
\end{proof}

\subsection{Proof of \hyref{proposition:sym-qre:global}[Proposition]}
\label{subsection:proposition:sym-qre:global}

\begin{proof}
    We first show that, in a sym-QRE, $P(\tilde x,\sigma)=1/2$.
    To this end, suppose $P(\tilde x,\sigma)>1/2$.
    Noting that, by definition, 
    $P(\tilde x,\sigma)=\frac{\theta^*(\sigma)-\tilde x+\epsilon}{2\epsilon}$,  
    $P(\tilde x,\sigma)>1/2$ thus implies that $\theta^*(\sigma)>\tilde x$.
    By {(3)} in \hyref{lemma:global:failure}[Lemma], 
    it must be that 
    $\theta^*(\sigma)\in(\tilde x-\epsilon,\tilde x+\epsilon)\Longleftrightarrow \tilde x\in(\theta^*(\sigma)-\epsilon,\theta^*(\sigma)+\epsilon)$,
    which, combined with the above observation delivers $\tilde x\in(\theta^*(\sigma)-\epsilon,\theta^*(\sigma))$.
    It is immediate from {(4)} in \hyref{lemma:global:failure}[Lemma] that 
    $\vert\Delta \bar u_{\tilde x+\delta}(\sigma)-\Delta \bar u_{\tilde x}(\sigma)\vert \geq \vert\Delta \bar u_{\tilde x-\delta}(\sigma)-\Delta \bar u_{\tilde x}(\sigma)\vert$
    for all $\delta\in[0,\epsilon]$. 
    But, by symmetry of $\sigma$, this implies that $\sigma(\tilde x+\delta)\geq \sigma(\tilde x-\delta)$ for all $\delta\in[0,\epsilon]$, which in turn, by definition of $\theta^*(\sigma)$, yields that $\theta^*(\sigma)\leq\tilde x$, a contradiction.
    A symmetric argument shows that it cannot be that $P(\tilde x,\sigma,\epsilon)<1/2$. 
    Hence, from {(2)} in \hyref{lemma:global:failure}[Lemma], $P(\tilde x,\sigma)= \frac{\tilde x - \underline \theta}{\overline \theta - \underline \theta}=1/2 \Longleftrightarrow \tilde x = (\overline \theta + \underline \theta)/2$.
    The result now follows directly from \hyref{theorem:sym-qre:characterization}[Theorem]. 

    We further note that, in this case, $\tilde x=\theta^*(\sigma)=(\overline \theta +\underline \theta)/2$.
    Then, $\Delta \bar u_{\tilde x +\delta}(\sigma)=\tilde x+\delta - \underline \theta -(\overline \theta-\underline \theta)P(\tilde x+\delta,\sigma)$.
    If $\delta\geq \epsilon$, $P(\tilde x +\delta,\sigma)=0$ and 
    $\Delta \bar u_{\tilde x +\delta}(\sigma)=\tilde x+\delta - \underline \theta=\delta + (\overline \theta - \underline \theta)/2$, whereas 
    $P(\tilde x -\delta,\sigma)=1$ and 
    $\Delta \bar u_{\tilde x -\delta}(\sigma)=\tilde x-\delta - \overline \theta=-\delta - (\overline \theta - \underline \theta)/2$.
    If $\delta\in (0,\epsilon)$, then 
    $P(\tilde x +\delta,\sigma)=0$ and 
    $\Delta \bar u_{\tilde x +\delta}(\sigma)
    =\tilde x+\delta - \underline \theta-(\overline \theta-\underline \theta)P(\tilde x + \delta, \sigma)
    =\tilde x+\delta - \underline \theta+(\overline \theta-\underline \theta)\delta/(2\epsilon)-(\overline \theta-\underline \theta)/2
    =
    \delta[1+(\overline \theta-\underline \theta)/(2\epsilon)]
    =
    \Delta \bar u_{\tilde x -\delta}(\sigma)
    $.
    In short, $\Delta \bar u_{\tilde x +\delta}(\sigma)=-\Delta \bar u_{\tilde x -\delta}(\sigma)$, for any $\delta \leq \min\{\tilde x,1-\tilde x\}$ and we conclude that symmetry
    requires $\sigma(\tilde x-\delta)=1-\sigma(\tilde x+\delta)$ for all $\delta\in[0,\min\{\tilde x,1-\tilde x\}]$. 
\end{proof}

\subsection{Proof of \hyref{proposition:rqre:compromise}[Proposition]}
\label{subsection:proposition:rqre:compromise}

\begin{proof}
    Fix a QRE $\sigma$. 
    Note that for the indifferent type $\tilde x$ has to satisfy 
    \begin{align*}
        G_\sigma(\tilde x)=B
        \Longleftrightarrow
        &\mathbb E[(1-\sigma(x'))1\{x'\leq \tilde x\}]=
        B\mathbb E[(1-\sigma(x'))1\{x'\leq \tilde x\}]
        +
        B\mathbb E[(1-\sigma(x'))1\{x'> \tilde x\}]
        \\
        \Longleftrightarrow
        &
        \mathbb E[(1-\sigma(x'))1\{x'\leq \tilde x\}](1-B)
        =
        B\mathbb E[(1-\sigma(x'))1\{x'> \tilde x\}]
        \\
        \Longleftrightarrow
        &
        \tilde x \mathbb E[(1-\sigma(x'))|{x'\leq \tilde x}](1-B)
        =
        (1-\tilde x)B\mathbb E[(1-\sigma(x'))|{x'> \tilde x}]
        \\
        \Longleftrightarrow
        &
        \tilde x 
        =
        \frac{B\mathbb E[(1-\sigma(x'))|{x'> \tilde x}]}{B\mathbb E[(1-\sigma(x'))|{x'> \tilde x}]+(1-B)\mathbb E[(1-\sigma(x'))|{x'\leq \tilde x}]}.
    \end{align*}
    From \hyref{theorem:rqre:characterization}[Theorem], $\sigma(\tilde x)=1/2$, and so $1>\mathbb E[(1-\sigma(x'))|{x'\leq \tilde x}]> 1/2 > \mathbb E[(1-\sigma(x'))|{x'> \tilde x}]>0$, implying 
    $0=\frac{0}{0+(1-B)}<\tilde x < \frac{B/2}{B/2+(1-B)/2}=B$.

    Now fix $\tilde x \in (0,B)$. 
    As before, let $\hat \sigma\in [0,1]^{\mathcal X}$ be such that $\hat \sigma(x):=\alpha$ for $x\in [0,\tilde x)$, $\hat \sigma(\tilde x)=1/2$, and $\hat \sigma(x):=\beta$ for $x\in (\tilde x,1]$, where $0<\alpha<1/2<\beta<1$.
    From $\Delta \bar u_{\tilde x}(\hat \sigma)=0$, we then have that $\tilde x = \frac {B(1-\beta)} {B(1-\beta)+(1-B)(1-\alpha)}$, with the right-hand side is continuous and increasing in $\alpha$ and $-\beta$, attaining $0$ when $\alpha<1/2, \,\beta=1$, and attaining $B$ when $\alpha=1/2, \,\beta=1/2$.
    For any $\alpha,\beta:0<\alpha<1/2<\beta<1$, one can then get a continuous and strictly increasing $\sigma \in (0,1)$ such that $\sigma(\tilde x)=\hat \sigma(\tilde x)=1/2$, $\mathbb E[\sigma(x')|x'\leq \tilde x]=\alpha$, and $\mathbb E[\sigma(x')|x'> \tilde x]=\beta$.
    Any such $\sigma$ will by \hyref{theorem:rqre:characterization}[Theorem] constitute a QRE.
\end{proof}

\subsection{Proof of \hyref{proposition:sym-qre:compromise}[Proposition]}
\label{subsection:proposition:sym-qre:compromise}

\begin{proof}
    Take $\hat \sigma(x)=\alpha$ for $x\in [0,\tilde x)$, $\hat \sigma(\tilde x)=1/2$, and $\hat \sigma(x)=1-\alpha$ for $x\in (\tilde x,1]$, where $0<\alpha<1/2$.
    From $\Delta \bar u_{\tilde x}(\hat \sigma)=0$, we then have that $\tilde x = \frac {B\alpha} {B\alpha+(1-B)(1-\alpha)}$, with the right-hand side is continuous and increasing in $\alpha$, attaining $0$ when $\alpha=0$, and attaining $B$ when $\alpha=1/2$.
    
    Moreover, since for any $x,x'$ such that $\Delta \bar u_{x}(\hat \sigma)=-\Delta \bar u_{x'}(\hat \sigma)$, $x<\tilde x<x'$ or $x>\tilde x>x'$, we get by construction $\hat \sigma(x)=1-\hat \sigma(x')$.
    For any $\alpha:0<\alpha<1/2$, one can then get a continuous and strictly increasing $\sigma \in (0,1)$ such that $\sigma(\tilde x)=\hat \sigma(\tilde x)=1/2$, $\mathbb E[\sigma(x')|x'\leq \tilde x]=\alpha$, and $\mathbb E[\sigma(x')|x'> \tilde x]=1-\alpha$.
    Any such $\sigma$ will by \hyref{theorem:sym-qre:characterization}[Theorem] constitute a sym-QRE.
\end{proof}

\subsection{Proof of \hyref{corollary:compromise}[Corollary]}
\label{subsection:corollary:compromise}

\begin{proof}
    For (1) observe that, for symmetric $\sigma$, we need that for $x,x'$ such that $\Delta \bar u_x(\sigma)=-\Delta \bar u_{x'}(\sigma)
    \Longleftrightarrow
    G_\sigma(x)=2B-G_\sigma(x') \Longrightarrow \sigma(x)=1-\sigma(x')$.
    Define $f:[0,\tilde x]\to [\tilde x,1]$ such that $f(x):={G_\sigma}^{-1}(2B-G_\sigma(x))$. 
    Then, $f$ is continuous and strictly decreasing, with $f(0)={G_\sigma}^{-1}(2B)$ and $f(\tilde x)={G_\sigma}^{-1}(B)=\tilde x$.
    Rewrite the above as $G_\sigma(\tilde x +\overline k)+G_\sigma(\tilde x -\underline k)=2B \Longleftrightarrow \sigma(\tilde x + \overline k)=1-\sigma(\tilde x - \underline k)$.
    Since $\sigma$ is continuous, $G_\sigma$ is continuously differentiable, with $\frac{\diff}{\diff k}G_\sigma(\tilde x + k)=\frac{1-\sigma(\tilde x + k)}{\mathbb E[1-\sigma(x'')]}$.
    Then, by the implicit function theorem, 
    $\frac{\diff \overline k}{\diff \underline k}=\frac{1-\sigma(\tilde x -\underline k)}{1-\sigma(\tilde x+\overline k)}=\frac{\sigma(\tilde x+\overline k)}{1-\sigma(\tilde x+\overline k)}$.
    Hence, $\frac{\diff \overline k}{\diff \underline k}>1$ for any $\underline k>0$, which implies that $\sigma(\tilde x+\underline k)<1-\sigma(\tilde x-\underline k)$ for any $\underline k\in[0,\tilde x]$.

    (2) follows from the fact that $f(0)={G_\sigma}^{-1}(2B)<{G_\sigma}^{-1}(1)=1$, and therefore, $1-\sigma(0)=\sigma(f(0))<\sigma(x)$ for $x>f(0)$. 
    Hence, setting $\overline x=f(0)$ gives the result.
\end{proof}

\section{Conditions (A1)-(A4) in applications}
\label{section:conditions}

In assessing \ref{assumption:a2}-\ref{assumption:a4}, we restrict $\sigma$ to $(0,1)^\mathcal X$.
This relaxation of \ref{assumption:a2}-\ref{assumption:a4} (requiring now that it only applies to such strategies) does not affect \hyref{theorem:rqre:characterization}[Theorems]-\hyref{theorem:recover}, since \ref{assumption:r1} implies that all QRE need to satisfy $\sigma\in (0,1)^\mathcal X$.

\subsection{Volunteer's dilemma}

We now show that \ref{assumption:a1}-\ref{assumption:a4} hold in the volunteer's dilemma.

\ref{assumption:a1}: 
Note that $\bar u_x^0(\sigma):=B-x$ and 
$\bar u_x^1(\sigma):=B(1 - \mathbb E[\sigma(x')])$.
It is immediate that both are jointly ($B$-Lipschitz) continuous in $(x,\sigma)$, as 
$|\bar u_x^0(\sigma)-\bar u_{x'}^0(\sigma')|=|x-x'|\leq B|x-x'|$ and 
$|\bar u_x^1(\sigma)-\bar u_{x'}^1(\sigma')|=B\left|\mathbb E[\sigma(x'')]-\mathbb E[\sigma'(x'')]\right|\leq B\|\sigma-\sigma'\|_{L^1}$.

\ref{assumption:a2}: Note that
$\bar u_{x'}^0(\sigma) - \bar u_{x}^0(\sigma)=x-x'$ 
and 
$\bar u_{x'}^1(\sigma) - \bar u_x^1(\sigma)=0$.
(A2) follows from the fact that, for any $\sigma \in \Sigma$ such that $\sigma(x) > \sigma(x')$ for some $x<x'$, we have that $\bar u_{x}^1(\sigma)=\bar u_{x'}^1(\sigma)$ and $\bar u_x^0(\sigma)>\bar u_{x'}^0(\sigma)$, and thus we satisfy (A2) with $\hat x = x$ and $\hat x'=x'$.

\ref{assumption:a3}: From the derivations above, (A3) is immediately obtained since, for any $x<x'$, 
$\bar u_{x'}^1(\sigma) - \bar u_x^1(\sigma)=0$ 
and 
$\bar u_{x'}^0(\sigma) - \bar u_{x}^0(\sigma)=x-x'<0$.

\ref{assumption:a4}: Since $\Delta \bar u_x(\sigma)=x-B\mathbb E[\sigma(x')]$, note that for $\sigma>1/2$ and $x=0$, $\Delta \bar u_x(\sigma)=-B\mathbb E[\sigma(x')]<0$,
while for $\sigma<1/2$ and $x=1$, $\Delta \bar u_x(\sigma)=1-B\mathbb E[\sigma(x')]>1-B/2>0$.

\subsection{Global games}

In the global games application, rather than establishing \ref{assumption:a1} and \ref{assumption:a2}, we have directly shown above that all QRE must be continuous and strictly increasing. 
With this, existence and characterization results go through unchanged.

We now show that (A3) and (A4) hold in the global game.

\ref{assumption:a3}: 
Let $\sigma$ be strictly increasing. 
This implies that $P(x,\sigma)$ is decreasing in $x$. 
Hence, $\bar u_x^1(\sigma)=x-c P(x,\sigma)$ is strictly increasing in $x$, whereas and $\bar u_x^0(\sigma)=k$ is constant.

\ref{assumption:a4}: 
As discussed, abstaining is strictly dominant for $x<\underline \theta$ and
attacking is strictly dominant for $x>\overline \theta$. 
Therefore, for all $\sigma$, $\Delta \bar u_x(\sigma)<0$ for sufficiently small $x$ and $\Delta \bar u_x(\sigma)>0$ for sufficiently large $x$.

\subsection{Compromise game}
In the compromise game, as discussed in Section \ref{subsection:compromise}, our characterization requires that we augment QRE with translation invariance.
This is because \ref{assumption:a2}-\ref{assumption:a4} are not satisfied given $\bar u_x^y(\sigma)$, but they hold when normalizing the payoffs to action $y=0$ to 0 and redefining the payoffs to action $y=1$ as the former difference in expected payoffs to action 1 and action 0, $\Delta\bar u_x(\sigma)$.

\ref{assumption:a1}: 
We have that $\bar u_x^1(\sigma)=\mathbb E[1\{x'\leq x\}]=x$, $\bar u_x^0(\sigma)=\mathbb E[\sigma(x')1\{x'\leq x\}]+B\mathbb E[1-\sigma(x')]$.
Note that $\|\bar u_x^1(\sigma)-\bar u_{x'}^1(\sigma')\|_{L^1}=|x-x'|$ and 
$
    \|\bar u_x^0(\sigma)-\bar u_{x'}^0(\sigma')\|_{L^1}
    \leq 
    B\|\sigma-\sigma'\|_{L^1}
    +
    \left|
        \int_0^{x}\sigma(\hat x)\diff \hat x
        -
        \int_0^{x}\sigma'(\hat x)\diff \hat x
    \right|
    +
    \left|
        \int_{x'}^{x}\sigma'(\hat x)\diff \hat x
    \right|
    \leq 
    (1+B)\|\sigma-\sigma'\|_{L^1}
    +
    |x-x'|
$.
Hence, $\bar u_x^y(\sigma)$ is Lipschitz-continuous in $(x,\sigma)$ for $y\in \{0,1\}$.

\ref{assumption:a2}: 
The expected payoff difference for type $x$ is 
$
    \Delta \bar u_x(\sigma)
    =
    (G_\sigma(x)-B)\mathbb E[1-\sigma(x'')] 
    = 
    \mathbb E[(1-\sigma(x''))1\{x''\leq x\}] - B\mathbb E[1-\sigma(x'')]
    =
    \int_{0}^{x}(1-\sigma(x''))\diff x''-B\mathbb E[1-\sigma(x'')] 
$.
Fix $\sigma\in (0,1)^\mathcal X$, with $\sigma(x) > \sigma(x')$ for some $x<x'$. 
Setting $\hat x=x$ and $\hat x'=x'$ yields $\Delta \bar u_{\hat x}(\sigma)-\Delta \bar u_{\hat x'}(\sigma)=-\int_{x}^{x'}(1-\sigma(x''))\diff x''<0$.

\ref{assumption:a3}:
From the above, we have that $\Delta \bar u_x(\sigma)$ is strictly increasing in $x$ for any $\sigma\in (0,1)^\mathcal X$.

\ref{assumption:a4}: 
The expected payoff difference for type $x$ is 
$
    \Delta \bar u_x(\sigma)
    =
    \int_{0}^{x}(1-\sigma(x''))\diff x''-B\mathbb E[1-\sigma(x'')].
$
For any $\sigma \in (0,1)$, $\Delta \bar u_0(\sigma)=-B\mathbb E[1-\sigma(x'')]<0$ and $\Delta \bar u_1(\sigma)=(1-B)\mathbb E[1-\sigma(x'')]>0$.

\section{Logit QRE and the set of indifferent types in the Volunteer's dilemma}
\label{section:volunteer:logit}

In logit QRE for a given value of $\lambda\in(0,\infty)$:
    $\sigma(x)={(1+\exp({\lambda(B\mathbb E[\sigma(x')]-x)}))}^{-1}$,
where $\mathbb E[\sigma(x')]$ denotes the ex-ante probability of abstaining. 
Taking expectations and rearranging delivers 
$$\frac{1}{\lambda}\ln \left(\frac{1+\exp(\lambda(B\mathbb E[\sigma(x')]-1))}{1+\exp(\lambda B \mathbb E[\sigma(x')])}\right)-\mathbb E[\sigma(x')]+1=0.$$
This implicitly defines the equilibrium values $\mathbb E[\sigma(x')]$ as a continuous function of $\lambda$, say $f(\lambda)$. 
Since $\mathbb E[\sigma(x')]\in (1/(B+1),1/2)$ for sym-QRE, the same must be true for logit QRE. 
By continuity of the dependence of $\mathbb E[\sigma(x')]$ on $\lambda$, it then suffices that we attain $1/2$ and $1/(B+1)$ as limits when $\lambda \downarrow 0$ and $\lambda \uparrow \infty$.
It is immediate that as $\lambda \downarrow 0$, all types will uniformly randomize, and so $\mathbb E[\sigma(x')]=1/2$.
In contrast, as $\lambda \uparrow \infty$, logit QRE converges to the unique Bayesian Nash equilibrium, satisfying $\mathbb E[\sigma^{BNE}(x')]=1/(B+1)$.

\end{document}